\documentclass[journal=jacsat,manuscript=article]{achemso}

\usepackage[version=3]{mhchem} % Formula subscripts using \ce{}
\usepackage{graphicx}
\usepackage{dcolumn}
\usepackage{bm}% bold math
\usepackage{qcircuit}
\usepackage{xcolor}
\usepackage{algorithm}
\usepackage{algpseudocode}
\usepackage{amsmath}
\usepackage{booktabs}
\usepackage{multirow}
\usepackage{hyperref}
\newcommand{\QW}[1]{\textcolor{black}{#1}}
\author{Qiaohong Wang}
\affiliation[University of Chicago]
{Pritzker School of Molecular Engineering, University of Chicago, Chicago, IL 60637, USA.}
\author{Ruhee D'Cunha}
\affiliation[University of Chicago]
{Department of Chemistry, Chicago Center for Theoretical Chemistry, University of Chicago, Chicago, IL 60637, USA.}
\author{Abhishek Mitra}%
\affiliation[University of Chicago]
{Department of Chemistry, Chicago Center for Theoretical Chemistry, University of Chicago, Chicago, IL 60637, USA.}
\author{Yuri Alexeev}
\affiliation{NVIDIA Corporation, 2788 San Tomas Express Way, Santa Clara, CA 95051}
\author{Stephen K. Gray}
\email{gray@anl.gov}
\affiliation{Center for Nanoscale Materials, Argonne National Laboratory, 9700 S. Cass Avenue, Lemont, IL 60439}
\author{Matthew Otten}
\email{mjotten@wisc.edu}
\affiliation{Department of Physics, University of Wisconsin – Madison, Madison, WI 53726, USA}
\author{Laura Gagliardi}
\email{lgagliardi@uchicago.edu}
\affiliation[University of Chicago]
{Department of Chemistry, Chicago Center for Theoretical Chemistry, University of Chicago, Chicago, IL 60637, USA.}
\title{Non-unitary Variational Quantum Eigensolver with the Localized Active Space Method and Cost Mitigation}


\begin{document}

\begin{tocentry}
\includegraphics[scale=0.328]{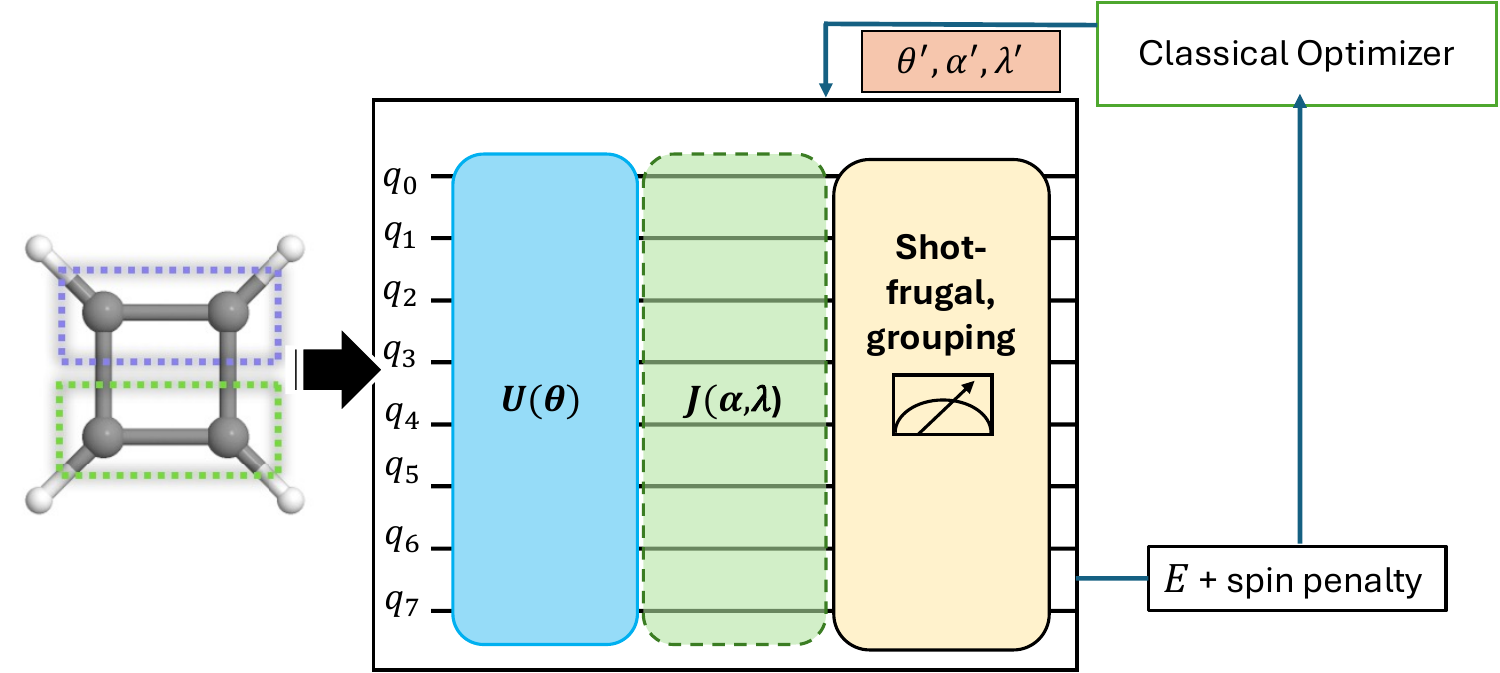}

%Inside the \texttt{tocentry} environment, the font used is Helvetica
%8\,pt, as required by \emph{Journal of the American Chemical

%The surrounding frame is 9\,cm by 3.5\,cm, which is the maximum
%permitted for  \emph{Journal of the American Chemical Society}
%graphical table of content entries. 

\end{tocentry}

\begin{abstract}
\QW{Accurately describing strongly correlated systems with affordable quantum resources remains a central challenge 
%in NISQ quantum algorithms for quantum chemistry.
for quantum chemistry applications on near and intermediate term quantum computers.
The localized active space self-consistent field (LASSCF) approximates the complete active space self-consistent field (CASSCF) by generating active space-based wave functions within specific fragments while treating interfragment correlation with mean-field approach, hence is computationally less expensive. Hardware-efficient ansatzes (HEA) offer affordable and shallower circuits, yet they often fail to capture the necessary correlation. Previously, Jastrow-factor-inspired non-unitary qubit operators were proposed to use with HEA for variational quantum eigensolver (VQE) calculations (so-called nuVQE), as they do not increase circuit depths and recover correlation beyond the mean-field level for Hartree-Fock initial states. Here, we explore running nuVQE with LASSCF as the initial state. The method, named LAS-nuVQE, is shown to recover interfragment correlations, reach chemical accuracy with a small number of gates ($<$70) in both H$_4$ and square cyclobutadiene (C$_4$H$_4$), and produces more accurate energetics than its HEA counterparts at all circuit depths. To further address the inherent symmetry-breaking in HEA, we implemented spin-constrained LAS-nuVQE to extend the capabilities of HEA further and show spin-pure results for square cyclobutadiene. We also mitigate the increased measurement overhead of nuVQE via Pauli grouping and shot-frugal sampling, reducing measurement costs by up to two orders of magnitude compared to ungrouped operator, and show that one can achieve better accuracy with a small number of shots (10$^{3-4}$) per one expectation value calculation compared to noiseless simulations with one or two orders of magnitude more shots. Finally, wall clock time estimates show that, with our measurement mitigation protocols, nuVQE becomes a cheaper and more accurate alternative than vanilla VQE with HEA. Taken together, these developments illustrate a practical pathway toward performing multireference chemical simulations with accuracy and affordable resources on today’s quantum hardware, achieving both accuracy and affordability in challenging correlated systems.}
\end{abstract}

\section{I. Introduction}
In recent decades, advances in quantum computing have led to new approaches to simulate many-body quantum systems; these show great promise owing to their polynomial scaling of quantum resources needed to solve otherwise exponentially complex quantum mechanical problems \cite{Intro1,QAforQCandQM,QCintheAGEofQC,QIAforCQM}. In the noisy, intermediate-scale quantum (NISQ) era \cite{Preskill2018quantumcomputingin}, hybrid algorithms such as the variational quantum eigensolver (VQE) are the most researched and developed choices for ground-state energy calculations of molecules \cite{Kandala2017,VQE,VQE_review,McClean_2016}. VQE approximates the lowest eigenvalue (i.e. ground-state) of a given operator (i.e. Hamiltonian) through the Rayleigh-Ritz variational principle. The algorithm iteratively adjusts parameters in a trial quantum circuit (a so-called ansatz) to minimize the expectation value of the Hamiltonian, which is calculated through measurements after each circuit's execution. The quantum state prepared with the optimal parameters hence represents the molecule's ground-state electronic wave function. As a result, the desired physical state is limited by the capability of the trial state and the affordability of the circuit depth. The search for an accurate and easy-to-prepare variational ansatz is an ongoing area of research. The variety of established variational ansatzes for VQE on quantum devices can be categorized into two families: 1) hardware-efficient ansatze (HEA) and 2) the chemically inspired unitary coupled cluster (UCC) ansatz, which was proposed to be used with VQE \cite{VQE}. Classically, UCC generates a non-terminating Baker-Campbell-Hausdorff series with a fixed excitation rank \cite{UCC_classical}, making it exponentially scaling for classical computers but only of polynomial complexity with Trotterization on a quantum computer \cite{poulin2014trotterstepsizerequired}. However, a large prefactor of the UCC and its expensive circuit depths (scales as $O(N^{4}_{q})$ with the $N_{q}$ number of qubits) limit its feasibility in NISQ devices, motivating the HEA as an alternative \cite{Kandala2017,Moll_2018}. The HEA (or heuristic ansatz) is designed to be efficiently implemented with gates natively available on quantum devices and is thus computationally advantageous. It usually leads to shallow circuit depths, which are necessary with hardware limitations such as coherence time. However, a potential flaw lies in the non-chemically motivated nature of the construction of HEA, as it does not guarantee the conservation of physical properties, such as spin and the number of particles \cite{ruhee_hardware}. 

Much progress has been made to reduce the complexity of UCC and develop alternative methods to provide an accurate and compact wave function ansatz. Several variations and derivatives of UCC include the qubit coupled-cluster ansatz \cite{qcc}, unitary paired CC ansatz \cite{GUCC}, iterative ansatzes like ADAPT-VQE and qubit ADAPT-VQE \cite{adapt,qADAPT}, selective schemes like the unitary selective coupled-cluster (USCC) ansatz \cite{Fedorov2022unitaryselective,lasuscc}. Another direction draws inspiration from the classical quantum Monte Carlo method, namely the unitary cluster Jastrow ansatz \cite{jastrow_decomp}. Efforts towards this direction include the local unitary cluster Jastrow ansatz \cite{lucj},
Jastrow quantum circuit \cite{PhysRevLett.123.130501},  and non-unitary VQE with linearized qubit Jastrow operators \cite{nuVQE}, which we explore in this work.  

A qualitatively good initial state is another key to the VQE algorithm. Multireference theory is indispensable to tackle complex electronic structure calculations, such as bond breaking, excited states, radicals, and transition-metal complexes~\cite{mcpdft}. Localized active space self-consistent field (LASSCF) \cite{LAS,vLASSCF}, also known as the cluster mean-field (cMF) method \cite{cmf}, approximates the paradigmatic multireference theory complete active space self-consistent field (CASSCF) \cite{casscf} by further partitioning the active space into fragments, ideally reducing the computational cost to linear with respect to the number of such fragments. The underlying assumption is that the electrons are strongly correlated in an active space that can be further separated into weakly interacting regions, resulting in a wave function in the form of an antisymmetrized product of fragment wave functions. LASSCF has been shown to predict spin-state energy gaps similar to those of CASSCF for bimetallic compounds~\cite{riddhish_spin}, describe accurate dissociations of two double bonds in bisdiazene~\cite{vLASSCF}, and produce a more consistent orbital evolution throughout the potential energy surface of sulfonium salts compared to CASSCF~\cite{wangsalts}. However, when electron correlation between fragments becomes important, the LASSCF approximation is too drastic, leading to inaccuracies. Restoring lost correlations classically using the variational \cite{mayhall2020}, perturbative \cite{perturb1,perturb2}, or CC approach \cite{corr_cc} is challenging for it usually requires building a many-body basis for each fragment. As a consequence, these approaches carry over the complexities inherent in multireference perturbation and CC theories.

The localized active space unitary coupled cluster (LAS-UCC) method successfully reintroduces correlations between LASSCF fragments with the aforementioned variational UCC ansatz with single and double excitations in a VQE scheme \cite{lasucc}, \QW{which allows accurate studies of multireference systems on quantum computers.} However, with hardware limitations in the NISQ era, running a LAS-UCC hardware calculation for relevant systems is impractical on current hardware. To reduce the cost, localized active space unitary selective coupled cluster (LAS-USCC) \cite{lasuscc} has been introduced, \QW{yet it might still be unrealistic to run while producing accurate results on near-term devices}. Recently, Benfenati et al. formulated non-unitary VQE (nuVQE) method with HEA~\cite{nuVQE} and showed that it produces more accurate energies for small molecules compared to just VQE with the same HEA at the same circuit depths. This approach could in principle provide another direction to \QW{recover interfragment correlations lost from LASSCF and study multireference systems with accuracy on quantum hardware} with much more affordable NISQ quantum resources. 

In this paper, \QW{we use LASSCF as the initial state to carry out nuVQE calculations with HEA. LASSCF provides a qualitatively good wave function as a starting point, and nuVQE ensures the affordability of quantum resources and improved accuracy compared to vanilla VQE with HEA}. The rest of the paper is laid out as follows: In Section II.\ref{sec:II}, we introduce the theory and formulations of LASSCF, nuVQE, and LAS-nuVQE. In Section III.\ref{sec:III}, we benchmark LAS-nuVQE performance on  H$_{4}$ and square cyclobutadiene. We also present a spin-constrained implementation of LAS-nuVQE. In addition, we showcase strategies to mitigate the measurement costs of nuVQE. Lastly, we provide resource estimation of wall clock time for one query, showing the scalability of our methods. In Section IV\ref{Sec:IV}, we provide some conclusions and discuss future directions.

\section{II. Theoretical background and computational methods\label{sec:II}}
\subsection{LASSCF}
Multireference methods like CASSCF and multireference configuration interaction (MRCI)~\cite{Buenker1974,Buenker1978} scale combinatorially with the number of electrons and orbitals, therefore presenting bottlenecks for large active spaces. Efforts to reduce the cost of CASSCF on classical computers include generalized active space (GAS) \cite{gasscf1,gasscf2,gasscf3,gasscf4}, restricted active space (RAS) \cite{rasscf,rasscf2}, density-matrix renormalization group (DMRG) \cite{CHAN2009149,dmrg2,dmrg3,dmrg4,dmrg5} self-consistent field \cite{dmrgscf,dmrgscf2} and LASSCF \cite{LAS,vLASSCF}. 

The LASSCF energy is optimized by variationally minimizing  
\begin{equation}
E_{\mathrm{LAS}}=\langle\mathrm{LAS}|\hat H| {\mathrm{LAS}}\rangle
\end{equation}
where $\hat H$ is the molecular Hamiltonian.
The wave function ansatz is
\begin{equation}
|\mathrm{LAS}\rangle = \bigwedge_{K} (\Psi_{A_K}) \wedge \Phi_D \label{LAS_equ}
\end{equation}
where $\Psi_{A_{K}}$ is a general many-body wave function to describe electrons occupying active orbitals of the $K$-th fragment, and $\Phi_{D} $ is a single determinant spanning the complement of the complete active space. The wedge operator $\bigwedge$ means that the active space wave function is an anti-symmetrized product of the fragment wave functions $\Psi_{A_{K}}$.

Classically solving LASSCF results in a fragmented orbital space and an effective Hamiltonian that is the sum of the local fragment Hamiltonians. Then, the fragment Hamiltonian goes through a fermion-to-spin transformation like the Jordan-Wigner transformation \cite{JW} to result in a qubit Hamiltonian, which is to prepare the fragment wave function with QPE circuits, direct initialization (DI), or VQE\cite{dcunha2023state}. We denote with
\begin{equation}
    |\Psi_{QLAS}\rangle = \bigwedge_{K} (\Psi_{A_k})
\end{equation} 
the active-space LASSCF wave function loaded onto a quantum device. The wedge operator $\bigwedge$ has the same meaning as the one in eq \eqref{LAS_equ}.

\subsection{Non-unitary VQE}
To fully capitalize on the short-circuit depths of HEA, nuVQE is one approach that increases the accuracy of the HEA without adding more gate operators to the circuit, ensuring the practicality of the algorithm in NISQ limitation. Generally, a nonunitary operator $\hat{O}(\vec{\lambda})$ and a unitary operator $\hat{U}(\vec{\theta})$ can be applied to an initial state $|\Psi_{0}\rangle$. The wave function can be expressed as
\begin{equation}
    |\Psi^{o}(\vec{\theta})\rangle = \hat{O}(\vec{\lambda})\hat{U}(\vec{\theta})|\Psi_{0}\rangle = \hat{O}(\vec{\lambda})|\Psi(\vec{\theta})\rangle ,
\end{equation}
where $\hat{U}(\vec{\theta})|\Psi_{0}\rangle $ is the HEA that evolves the initial states with some unitary gates, resulting in the trial wave function $|\Psi(\vec{\theta})\rangle$. After applying operator $\hat{O}(\vec{\lambda})$, the energy of the wave function is given by
\begin{equation}\label{equ:normal}
E=\frac{\left\langle\Psi^{o}(\vec{\theta})\left| \hat{H} \right| \Psi^{o}(\vec{\theta})\right\rangle}{\langle\Psi^{o}(\vec{\theta})| \Psi^{o}(\vec{\theta})\rangle},
\end{equation}
which can be re-expressed as
\begin{equation} \label{equ:oho}
E=\frac{\left\langle\Psi(\vec{\theta})\left|\hat{O}^{\dagger}(\vec{\lambda}) \hat{H} \hat{O}(\vec{\lambda})\right| \Psi(\vec{\theta})\right\rangle}{\left\langle\Psi(\vec{\theta})\left|\hat{O}^{\dagger}(\vec{\lambda}) \hat{O}(\vec{\lambda})\right| \Psi(\vec{\theta})\right\rangle}.
\end{equation}
Noticeably, implementing non-unitary operations as eq \eqref{equ:normal} is challenging. One can implement it on a gate-based quantum computer by adding ancillary qubits, as described in~\cite{PRXQuantum.2.010342}. Eq \eqref{equ:oho}, on the other hand, keeps the shallow HEA circuit and allows one to calculate the expectation values of $\hat{O}^{\dagger} \hat{H} \hat{O}$ and $\hat{O}^{\dagger}\hat{O}$ while optimizing parameters $\vec{\lambda}$ and $\vec{\theta}$. This formulation can introduce chemically-inspired non-unitary operators while mantaining the same circuit depth as a vanilla VQE with HEA, with the only extra price of the additional measurement of the wave function, which is not normalized.

Many potential forms of non-unitary operators exist and the Jastrow factors~\cite{jastrow} used in Quantum Monte Carlo methods~\cite{qmc_review} are classically formulated to introduce accurate and explicitly correlated wave functions beyond the mean-field level. Recent work in the field has constructed qubit operators inspired by one-body and two-body Jastrow factors \cite{nuVQE} in the form of
\begin{equation}
    J = J_{1} + J_{2}
\end{equation}
where
\begin{equation}
J_1=\exp \left[-\sum_{i=1}^N \alpha_i Z_i\right],
J_2=\exp\left[-\sum_{i<j=1}^N \lambda_{i,j} Z_iZ_j\right],
\end{equation}
where $\alpha_i$ and $\lambda_{i,j}$ are real coefficients that are sampled from a uniform distribution, and $Z_i = \otimes_j\left(I_2\right)_{j \neq i}(Z)_{j=i}$, where $i,j$ are qubit indices. 
In addition, the authors approximate the operator by linearizing it to further reduce the terms needed for the measurement in the form of
\begin{equation}
    J(\vec{\alpha},\vec{\lambda})= 1 - \sum_{i=1}^N \alpha_i Z_i - \sum_{i<j=1}^N \lambda_{i,j} Z_iZ_j ,
\end{equation}
\subsection{LAS-nuVQE}
\begin{figure*}
    \centering \includegraphics[width=\linewidth]{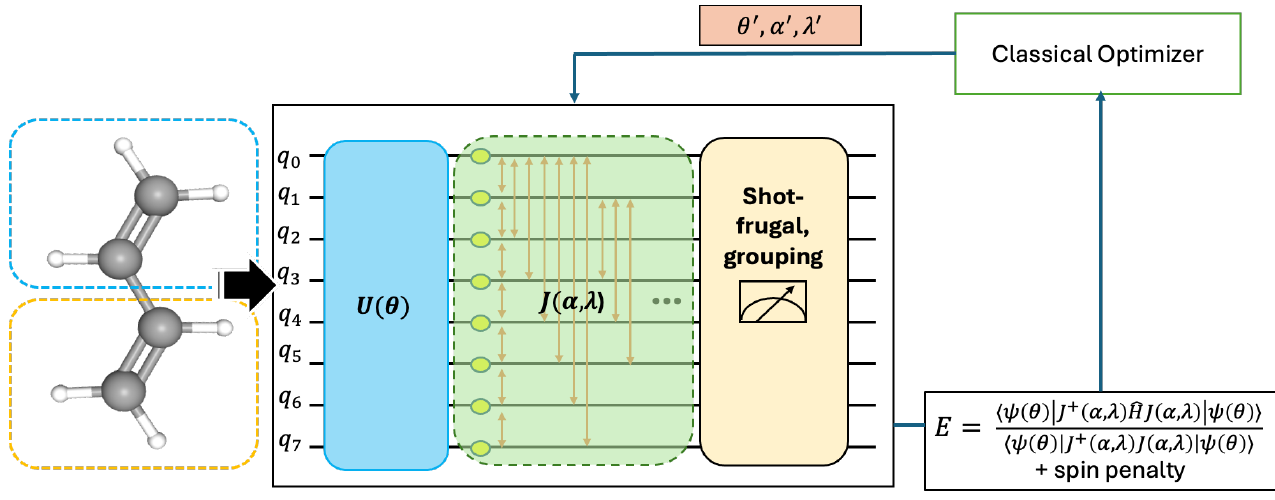}   \caption{Flowchart describing the LAS-nuVQE algorithm. \textit{Left to right:} An example molecular system solved with LASSCF, where two boxes indicate fragmentation. We then prepare a parameterized HEA with LASSCF as the initial state with the DI method. One-body and two-body Jastrow parameters are classically generated, the dashed lines indicate that the Jastrow parameters do not operate on the circuit. The yellow measurement block illustrates the grouping and shot-frugal techniques one can implement to mitigate measurement costs for the energy evaluation. Lastly, the classical optimizer gives the next set of parameters to optimize the energy expression until convergence variationally.  }
    \label{fig:flowchart}
\end{figure*}
\begin{algorithm}
\caption{Localized active space non-unitary variational quantum eigensolver}
\label{lasnuvqe_alg}
\begin{algorithmic}[1]
    \State Prepare the LASSCF wave function on the quantum computer with DI, construct hardware-efficient ansatz
    \State Generate a set of parameters defining $J_1$ and $J_2$, and calculate Pauli strings for $\hat{J}^{\dagger}\hat{H}\hat{J}$ and $\hat{J}^{\dagger}\hat{J}$.
    \State Minimize the energy expression in eq~\eqref{lasnuvqe_energy} by variationally optimizing the  $\vec{\theta}, \vec{\alpha}, \vec{\lambda}$ parameters using a classical optimizer.
\end{algorithmic}
\end{algorithm}
The LAS-nuVQE algorithm is illustrated in Fig.\ref{fig:flowchart} and explained step-by-step in Algorithm \ref{lasnuvqe_alg}. The extra number of classical parameters introduced by the Jastrow operators scales as $\frac{1}{2}N_{q}(N_{q}+1)$, where $N_{q}$ is the number of qubits. The cost function is defined as
\begin{equation}\label{lasnuvqe_energy}
E=\frac{\left\langle\Psi_{QLAS}(\vec{\theta})\left|\hat{J}^{\dagger}(\vec{\alpha},\vec{\lambda}) \hat{H}_{QLAS} \hat{J}(\vec{\alpha},\vec{\lambda})\right| \Psi_{QLAS}(\vec{\theta})\right\rangle}{\left\langle\Psi_{QLAS}(\vec{\theta})\left|\hat{J}^{\dagger}(\vec{\alpha},\vec{\lambda}) \hat{J}(\vec{\alpha},\vec{\lambda})\right| \Psi_{QLAS}(\vec{\theta})\right\rangle},
\end{equation}
where $\Psi_{QLAS}(\vec{\theta})$ is the HEA with the LASSCF wave function as the initial state (i.e. step 1 in Algorithm 1), and $\hat{H}_{QLAS}$ is the Hamiltonian of the active space in the qubit basis. In addition to the increase in parameters needed to be optimized compared to HEA, the measurements naively scale $O(N_{q}^{8})$ compared to $O(N_{q}^{4})$. The increased measurement costs, however, can be further reduced with operator grouping and shot-frugal techniques, as described in Section III D. 
\subsection{Computational details}
The ground state energies of the following chemical systems are computed: H$_4$ at two different geometries and square cyclobutadiene, where the simulations require qubits counts of 8 respectively.
\begin{figure}[h!]
\includegraphics[scale=0.6]{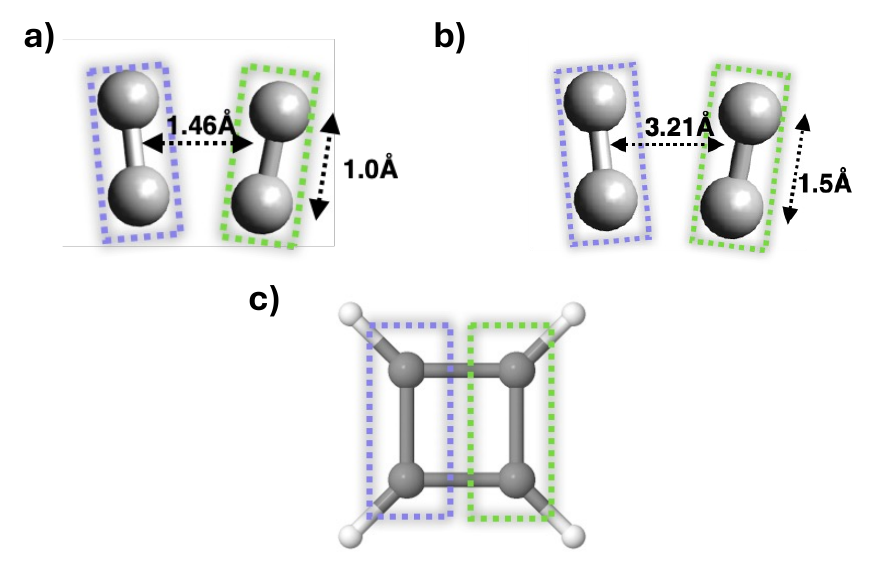}% Here is how to import EPS art
\caption{Systems studied in this work with purple and green boxes indicating the LASSCF fragments: \textbf{a)} two hydrogen molecules (intramolecular distance = 1.0 \r{A}) separated by 1.46 \r{A} (Geometry 1). \textbf{b)} two hydrogen molecules (intramolecular distance of  1.5  \r{A}) separated by 3.21 \r{A} (Geometry 2). \textbf{c)} square cyclobutadiene where all C-H bonds are 1.069 \r{A} and C-C bonds are 1.456 \r{A}. \label{fig:2systems}}
\end{figure}
All molecular simulations with VQE and nuVQE (Section III.\ref{sec:III} A. to C.) are noiseless and carried out with DI state preparation with Qiskit (version 0.46) \cite{qiskit2024}, Aer state vector simulator (qiskit-aer version 0.11.1) and Limited-memory BFGS Bound(L-BFGS-B) optimizer \cite{lbfgsb} with \textit{scipy}. The initial values of $\alpha_i$ and $\lambda_{i,j}$ are sampled from a distribution, $\text{Uniform}(-\epsilon,\epsilon)$, where $\epsilon = 0.1$ in our calculations. LASSCF calculations are performed with the $mrh$~\cite{mrh} package, and HF related calculations with \textit{pyscf} \cite{Pyscf}. To illustrate Pauli grouping and shot-frugal techniques to optimize the algorithm, we performed noiseless QASM simulations with the QASM simulator from the Qiskit BasicAer module, as described in Section \ref{sec:III} D. The corresponding basis sets, active spaces, and simulation details for each example are introduced in the respective sections.

\section{III. Results and Discussions\label{sec:III}}
\subsection{Hardware-efficient ansatz comparison}
Many forms of HEA exist due to the use of native quantum gates like rotation-y (ry) gates and CNOT gates. Before discussing the nuVQE algorithm, we benchmark the performance and effectiveness of using the HEA as the trial wave function for VQE calculations using various initial states. 
\begin{figure}[h!]
\includegraphics[scale=0.75]{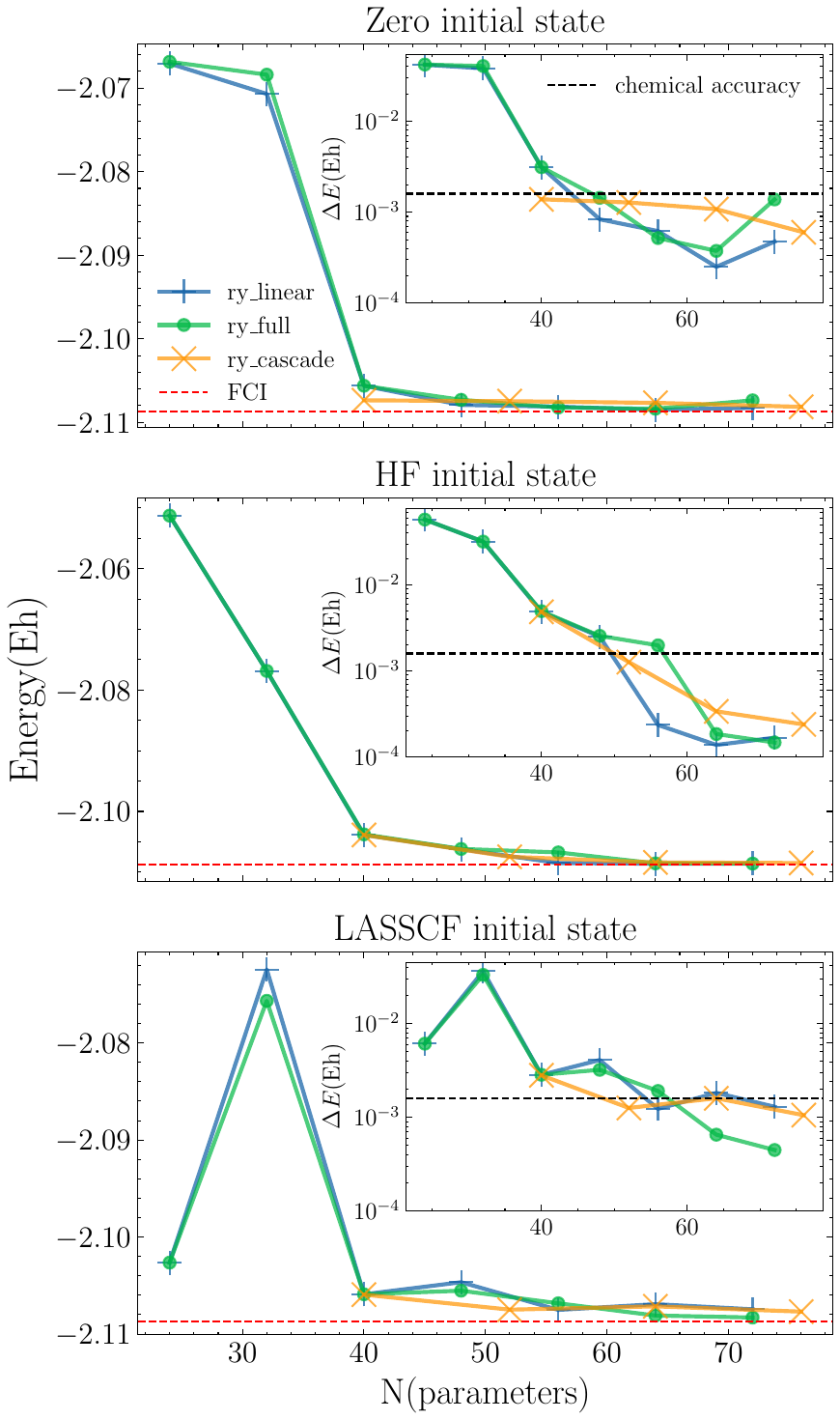}% Here is how to import EPS art
\caption{Performance benchmark of three hardware-efficient ansatzes: linear, full and cascade connectivity on system H$_4$ (Geometry 1) with minimal STO-3G basis set and increased number of layers (2-9 for ry\_linear and ry\_full, 2-5 for ry\_cascade). Three different initial states: 0, Hartree-Fock (HF), and LASSCF are also being used to further compare the HEA. \label{fig:hardware_ansatz_compare}}
\end{figure}
We compare the performance of three widely-used ansatzes with $Ry$ rotation gates and different connectivity, namely linear, full, and cascade (see Appendix A\ref{sec:appendixHEA}), for the H$_4$ system with Geometry 1 and minimal STO-3G basis set (8 qubits used). We also compare three initial states: all zero parameters, Hartree-Fock (HF), and LASSCF. For LASSCF, we choose an active space of (2e,2o) for each fragment (H$_2$). We carry out noiseless VQE calculations with the statevector simulator, and the data plotted are taken from the lowest energy calculation from 1,000 independent runs. 

Since each additional layer adds a different number of parameters to $Ry$\_cascade compared to $Ry$\_linear and $Ry$\_full, we use the number of variational parameters in the ansatzes as x-axis and each point represents an additional layer for the respective ansatz. From Figure \ref{fig:hardware_ansatz_compare}, we first observe that all but one energy values are above the chemical accuracy before 40 parameters, which corresponds to four repetitions of the linear and full connectivity, and one layer of the cascade connectivity. HEA thus fails to report accurate energies with short circuit depths ($<4$ repetitions).  When we increase the circuit depth by adding more layers, HEA agrees with FCI within chemical accuracy. However, the energy is not decreasing with more repetitions, although there are more variational parameters in the algorithm. Noticeably, the LASSCF initial state with two repetitions (24 variational parameters) outperforms the other two initial states and has significantly lower energy errors. We also observe that there is not one connectivity that substantially outperforms the others. Hence, we utilize $Ry\_linear$ as our choice of HEA for the rest of the results. 

\subsection{Single-point calculations}
\subsubsection{H$_{4}$ system}
For the initial testing of LAS-nuVQE, we explored the two H$_4$ system illustrated in Figure \ref{fig:2systems} with a minimal STO-3G basis set and a Jordan-Wigner transformation \cite{JW}. An active space of (2e, 2o) for each fragment is used corresponding to a total number of 8 qubits. We run 1,000 independent VQE calculations and 100 independent nuVQE calculations at all circuit repetitions (2 layers to 6 layers) with optimizer L-BFGS-B at each geometry. The plotted data corresponds to the lowest energy result. The total number of classical parameters to be optimized is 60 (24 for HEA and 36 from Jastrow operators). 
\begin{figure}[h!]
\includegraphics[scale=0.7]{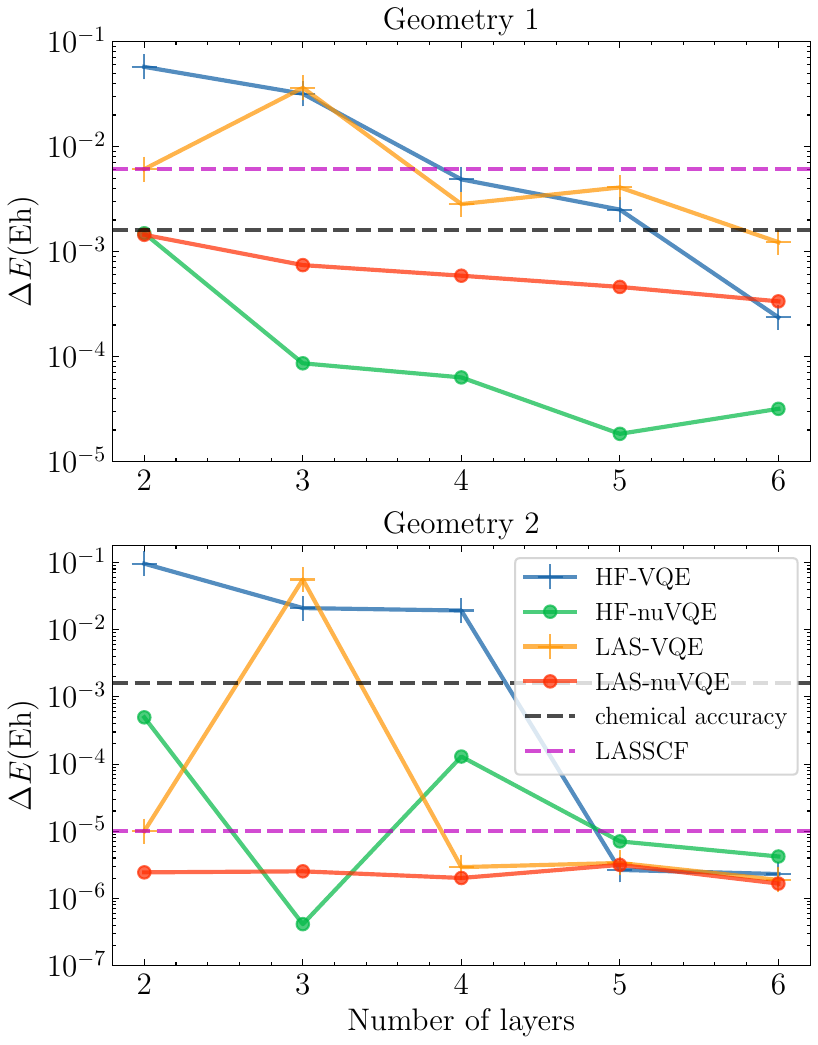}
\caption{VQE and nuVQE results with HF and LASSCF initial states for two geometries of H$_4$ and minimal basis set STO-3G. x-axis is the number of circuit repetitions for HEA, and y-axis is the log scale of energy difference from full configuration interaction (FCI) reference. \label{fig:h4_2geom}}
\end{figure}

\QW{As Figure \ref{fig:h4_2geom} shows, at geometry 1, LASSCF is above chemical accuracy, indicating an approximation that is too drastic. LAS-nuVQE achieves chemical accuracy with only 2 layers of HEA, which translates to 24 single-qubit gates (SQGs) and 21 CNOT gates ($<50$ total number of gates). It illustrates that LAS-nuVQE can recover interfragment correlations and gives a more accurate energy. In addition, LAS-nuVQE is closer to the FCI reference at all circuit repetitions than LAS-VQE. Similar behavior is observed with the HF initial state as well. For geometry 2, LASSCF is a good approximation, indicated by the energy well below chemical accuracy. nuVQE is still able to improve LASSCF energies. Again, LAS-nuVQE outperforms the LAS-VQE counterparts at all circuit layers. This means that at the same number of HEA layers, LAS-nuVQE is more accurate than LAS-VQE.} These test results encouraged us to explore more realistic systems. 
\subsubsection{Square Cyclobutadiene}
Cyclobutadiene is an anti-aromatic and highly strained molecule. Its $1^{1}B_{1g}$ ground state has two degenerate singly occupied frontier orbitals and represents a challenge for single configuration methods \cite{cyclo1}. We calculated this open-shell singlet ground-state at the square geometry, where the four carbon-carbon bonds are 1.456 \r{A}, and the carbon-hydrogen bonds 1.069 \r{A}. The chosen active space is (4e, 4o), which is further broken down into two (2e, 2o) fragments corresponding to a total number of 8 qubits. The cc-pvdz basis set \cite{ccpvdz} was used. The results presented are taken from 1,000 independent VQE calculations and 100 independent nuVQE at all circuit repetitions (2 layers to 6 layers) with optimizer L-BFGS-B. 
\begin{figure}[h!]
    \centering
    \includegraphics[scale=0.7]{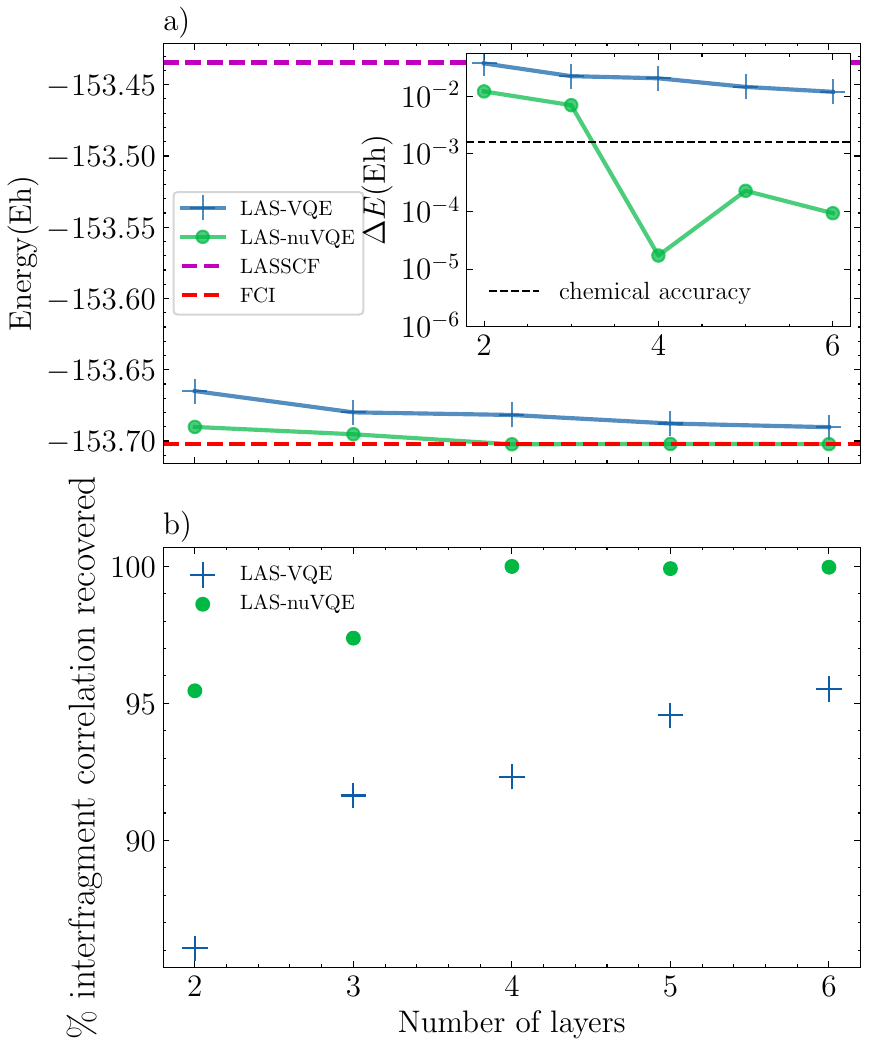}
    \caption{VQE and nuVQE comparion for square cyclobutadiene. \textbf{a)} plots the energy vs. the number of HEA layers, with inset's y-axis being the log scale of the energy difference from CASCI reference. The bottom panel \textbf{b)} calculates the \% of interfragment correlation recovered from LASSCF.}
    \label{fig:cyclobu}
\end{figure}

As Figure \ref{fig:cyclobu} (a) shows, LAS-nuVQE reaches chemical accuracy with 4 layers, which translates to 40 SQGs, 28 CNOT gates ($<70$ total number of gates) and a total number of classical parameters of 76 (40 from HEA and 36 from Jastrow operators). The LAS-nuVQE energy is below the LAS-VQE value at all circuit repetitions. Noticeably, LAS-VQE never reaches chemical accuracy even with 6 repetitions, suggesting that one has to increase the circuit depth drastically to obtain energies comparable to the LAS-nuVQE results. In addition, we calculated the \% of missing interfragment correlations recovered from LAS-nuVQE and LAS-VQE (Figure \ref{fig:cyclobu}(b)) as a function of the number of layers for both methods as
\begin{equation}
   p_0 = \frac{E_{LASSCF}-E_{LAS-nuVQE}}{E_{LASSCF}-E_{CASCI}}.
\end{equation}

We notice that nuVQE recovers 95\%$+$ of the interfragment correlation with 2 layers, and above 99.9\% for 4 and above repetitions. 
\subsection{Spin-constrained calculations}
One of the pitfalls of HEA is that it cannot preserve the total spin ($\hat{S^{2}}$) of the wave function, where the total spin operator can be measured with $\hat{S^{2}} = \hat{S_{+}}\hat{S_{-}} + \hat{S_{z}^{2}} +\hbar\hat{S_{z}}$. In the case of square cyclobutadiene reported in Section III\ref{sec:III}, the total spin of the wave function should be 0, while we observe spin contaminations ranging from 0.67 to 1.52 in all calculations as reported in Table \ref{tab:cyclobutadiene_un}. 
\begin{table}[h!]
\renewcommand{\arraystretch}{1.2}
\setlength{\tabcolsep}{7.5pt}
\caption{Unconstrained LAS-VQE and LAS-nuVQE results for cyclobutadiene reported in  Section \ref{sec:III} B. Energy (in Hartree) corresponds to the points plotted in Figure 
\ref{fig:cyclobu}. VQE is taken as the lowest energy result from 1000 independent runs, nuVQE is taken from 100 independent runs.\label{tab:cyclobutadiene_un}}
\begin{tabular}{|l|l|l|l|}
\hline
Method & Layer & Energy & Spin($\hat{S^{2}}$)\\ \hline
\multirow{5}{*}{\begin{tabular}[c]{@{}l@{}}Unconstrained\\ LAS-VQE\end{tabular}} 
 & 2 & -153.66495918 & 1.23278361 \\ 
 & 3 & -153.67988680& 1.20112778\\ 
 & 4 & -153.68172310 & 1.14305232\\  
 & 5 & -153.68772309 & 0.9311722\\
 & 6 & -153.69030785& 0.79768462 \\\hline
\multirow{5}{*}{\begin{tabular}[c]{@{}l@{}}Unconstrained\\ LAS-nuVQE\end{tabular}} 
 & 2 & -153.69012437 & 1.15427599 \\ 
 & 3 & -153.69526760 & 0.67047731 \\ 
 & 4 &  -153.70227735& 0.99937762\\  
 & 5 &  -153.70206538 & 1.51563038 \\
 & 6 &  -153.70220104 & 1.06992523 \\ \hline
\end{tabular}
\end{table}
To perform meaningful chemical calculations with LAS-nuVQE, we implemented a spin penalty term in the energy optimization cost function \cite{penalty_spin},
\begin{equation}   E(\boldsymbol{\theta}):=L(\boldsymbol{\theta})+\mu_C\left\langle\Psi_{QLAS}(\boldsymbol{\theta})\left|(\hat{C}-c)^2\right| \Psi_{QLAS}(\boldsymbol{\theta})\right\rangle
\end{equation}
where $L(\boldsymbol{\theta)}$ is the energy calculated using eq \eqref{lasnuvqe_energy}, $\mu_C = 1$~\cite{penalty_spin}, $\hat{C}$ is the total spin operator in this specific case, and $c$ is the ideal spin state.

%Noticeably, adding the spin constraint complicates the optimization landscape and increases the time for VQE runs in general. As a result, we only report below the energy values from 100 independent VQE and nuVQE runs for square cyclobutadiene as a capability demonstration that one can obtain spin-pure results.
%\subsubsection{Square cyclobutadiene}

\begin{table}[h!]
\renewcommand{\arraystretch}{1.2}
\setlength{\tabcolsep}{7.5pt}
\caption{Spin-constrained LAS-VQE and LAS-nuVQE results for cyclobutadiene reported. Energy unit is in Hartree, and both VQE and nuVQE are taken as the lowest result among 50 calculations\label{tab:cyclobutadiene_c}.}
%VQE: 4 layer(70/100),5 layer:(60/100),6 layer: (50/100); nuVQE: 2 layer:(61/100),3 layer (50/100),4 layer(44/100),5 layer (39/100),6 layer(35/100)}
\begin{tabular}{|l|l|l|l|}
\hline
Method & Layer & Energy & Spin($\hat{S^{2}}$)\\ \hline
\multirow{5}{*}{\begin{tabular}[c]{@{}l@{}}Constrained\\ LAS-VQE\end{tabular}}
 & 2 & -153.64195397& 1.15821369e-09\\ 
 & 3 & -153.64195398& 2.34582602e-09 \\ 
 & 4 & \textcolor{black}{-153.66371492}&\textcolor{black}{9.58401288e-05}  \\  
 & 5 &\textcolor{black}{-153.66427448}  &\textcolor{black}{3.31614262e-05} \\ 
 & 6 & \textcolor{black}{-153.66417350} & \textcolor{black}{0.00056101} \\\hline
 \multirow{5}{*}{\begin{tabular}[c]{@{}l@{}}Constrained\\ LAS-nuVQE\end{tabular}} & 2 & \textcolor{black}{-153.68087791} &\textcolor{black}{5.94048096e-06}  \\ 
 & 3 & \textcolor{black}{-153.67765748} & \textcolor{black}{7.17730272e-05} \\ 
 & 4 & \textcolor{black}{-153.67582877} &\textcolor{black}{0.00024698}  \\  
 & 5 & \textcolor{black}{-153.69621423} &\textcolor{black}{0.00043989}  \\ 
 & 6 & \textcolor{black}{-153.66854958} &\textcolor{black}{0.00023403} \\\hline
\end{tabular}
\end{table}
As Table \ref{tab:cyclobutadiene_c} shows, we have constrained the HEA to the correct spin state. For validation of the spin conservation, we run multiple independent runs and show the lowest energy results taken from 50 independent runs from both methods.
%Noticeably, adding the extra term complicates the optimization landscape and increases the optimization time. As a result, each independent VQE run time is comparable to nuVQE and we compare results taken only from \QW{~50/100} independent runs from both methods. 
We expect that with more optimization runs, we could achieve lower energies. % is also why the constrained VQE values seem to be higher than the previous unconstrained results, which are taken from 1000 independent runs at each layer. 
Our results again have shown that at the same number of layers, LAS-nuVQE always improves on LAS-VQE results. 

\subsection{Mitigating measurement costs}
\QW{So far, we have shown that LAS-nuVQE is effective in recovering correlations from LASSCF with small systems and is affordable to run with quantum resources available today. However, to tackle larger chemical systems in the future, we need to optimize the LAS-nuVQE algorithm, namely the measurement overhead, to make LAS-nuVQE affordable.}
The number of Pauli strings for the Hamiltonian operator scales naively as $O(N_{q}^4)$. The modified Pauli strings  $J(\vec{\alpha},\vec{\lambda})^{\dagger}HJ(\vec{\alpha},\vec{\lambda})$ could potentially scale up to $O(N_{q}^8)$ and the cost of $J(\vec{\alpha},\vec{\lambda})^{\dagger}J(\vec{\alpha},\vec{\lambda})$ is one due to only Pauli $Z$ and $I$ present in the formulation and that the strings all commute. Measurement optimization and efficiency are highly relevant in the NISQ era. To overcome the measurement costs of the nuVQE algorithm, we explored Pauli grouping and shot-frugal techniques to illustrate that the method could be further optimized. Other methods  to mitigate measurement costs not investigated in this work include Majorana classical shadow \cite{shadow1,shadow2}, low-rank decomposition \cite{lowrank1,lowrank2,lowrank3}, and fluid Fermionic fragments (F$^{3}$) \cite{Choi2023fluidfermionic}.
\subsubsection{Pauli grouping}
For any quantum operators $\hat{O}$, after mapping to a qubit operator $\hat{O} = \sum_i c_i\hat{P}_i$, the expectation value of that operator can be expressed as the weighted sum of the measurements of Pauli strings with their respective coefficients
\begin{align}
\label{expectvalue}
\langle\hat{O}\rangle & =\langle\Psi|\hat{O}| \Psi\rangle=\langle\Psi|\sum_i c_i \hat{P}_i| \Psi\rangle \notag \\
& =\sum c_i\langle\Psi|\hat{P}_i| \Psi\rangle=\sum c_i\langle\hat{P}_i\rangle.
\end{align}
\begin{table*}[h!]
\renewcommand{\arraystretch}{1.2}
\setlength{\tabcolsep}{2.5 pt}
  \caption{Comparison of the number of groups of Pauli operator after qubit-wise commuting (QWC) and fully commuting (FC) techniques for systems of H$_2$,H$_4$, and cyclobutadiene (C$_4$H$_4$) with three different mapping schemes. }
   \begin{tabular}{llccccccccc}
\hline\hline
       \multicolumn{2}{c}{\multirow{2}{*}{}} & \multicolumn{3}{c}{H$_2$} & \multicolumn{3}{c}{H$_4$} & \multicolumn{3}{c}{C$_4$H$_4$}\\
\cmidrule(lr){3-5}\cmidrule(lr){6-8}\cmidrule(lr){9-11}
       \multicolumn{1}{c}{} & Pauli operator & Original & QWC  & FC & Original & QWC  & FC & Original & QWC  & FC \\ \hline 
\multirow{3}{*}{Jordan-Wigner} & Hamiltonian & 185 & 46 & 9 & 361 & 73 & 19 &249&44&12\\
 & $JHJ$ & 3147 & 185 & 29 & 5879 & 349 & 59&3147&44&12\\
 & $JJ$ & 163 & 1 & 1 & 163 & 1 & 1&163&1&1 \\ \hline
 \multirow{3}{*}{Bravyi-Kitaev} & Hamiltonian & 185 & 34 & 9 & 361 & 70 & 19&361&70&23 \\
 & $JHJ$ & 3472 & 137 & 27 & 6448 & 269 & 63 &6442&70&23\\
 & $JJ$ & 163 & 1 & 1 & 163 & 1 & 1&163&1&1\\ \hline
 \multirow{3}{*}{Parity} & Hamiltonian & 185 & 34 & 9 & 361 & 73 & 19 &217&36&13\\
 & $JHJ$ & 3376 & 143 & 27 & 6350 & 269 & 63&3372&143&32\\
 & $JJ$ & 163 & 1 & 1 & 163 & 1 & 1 &163&1&1 \\
\hline
\end{tabular}
\label{pg-table}%
\end{table*}%

For diagonal Pauli operators, the expectation value in the computational basis ($I$ and $Z$) can be directly measured. For nondiagonal operators such as the $X$ and $Y$ Pauli operators, the corresponding matrices can go through a basis change to be further diagonalized. Pauli grouping is hence a commonly-used method to reduce the necessary number of evaluations on a quantum hardware. By gathering Pauli strings into groups of simultaneously diagonalizable strings (i.e. share the same eigenbasis), the expectation value for each group's Pauli strings can be determined with a single measurement with a classical post-processing step \cite{Kandala2017,McClean_2016}.
Two common partitions include qubit-wise commutativity (QWC) \cite{qwc,Kandala2017} and full commutativity (FC) \cite{fc}. In QWC, two Pauli strings commute if the matrices commute at each index, whereas in FC, the commutation rule is applied to the whole operator. 

Finding the optimal Pauli partition is the same as solving a graph partition problem known as the minimum clique cover problem, which is NP hard \cite{qwc}. We hence only focus on applying QWC and FC techniques and illustrate the savings in Table \ref{pg-table} for systems H$_2$, H$_4$, and C$_4$H$_4$. The H$_2$ and H$_4$ Hamiltonians encompass the full system, whereas C$_4$H$_4$ only includes the active space. We investigated three mapping schemes, Jordan-Wigner \cite{JW}, Bravyi-Kitaev \cite{BRAVYI2002210}, and Parity \cite{parity} for three operators: Hamiltonian ($H$), $J(\vec{\alpha},\vec{\lambda})^{\dagger}HJ(\vec{\alpha},\vec{\lambda})$, and $J(\vec{\alpha},\vec{\lambda})^{\dagger}J(\vec{\alpha},\vec{\lambda})$ respectivity. Across three mapping schemes, the measuring costs of the Hamiltonian operator is reduced by a factor of $\sim 4-21$ and $J(\vec{\alpha},\vec{\lambda})^{\dagger}HJ(\vec{\alpha},\vec{\lambda})$ is reduced by a factor of  $\sim 17-280$, which brings the total increased cost down to as littles as a factor of 3. It shows that despite nuVQE introducing considerable measurement overhead, it can be efficiently mitigated with Pauli grouping techniques, which have already been implemented in various hardware platforms.

\subsubsection{Shot-frugal techniques}
As previously discussed, the expectation value of a quantum operator is usually taken as the weighted sum of measured Pauli strings according to eq \eqref{expectvalue}. 
The terms with a larger coefficient `contribute' more to the final numerical expression and their accuracy might be more important. When measuring an expectation value with a limited number of shots, various strategies—often referred to as shot-frugal techniques—have been developed to optimize shot distribution and achieve high accuracy while minimizing resource usage \cite{arrasmith2020operatorsamplingshotfrugaloptimization,Menickelly2023latency}.  In this work, we leverage such strategies to reduce the computational cost associated with evaluating  $J(\vec{\alpha},\vec{\lambda})^{\dagger}HJ(\vec{\alpha},\vec{\lambda})$. 

\begin{figure}[h!]
\includegraphics[scale=0.65]{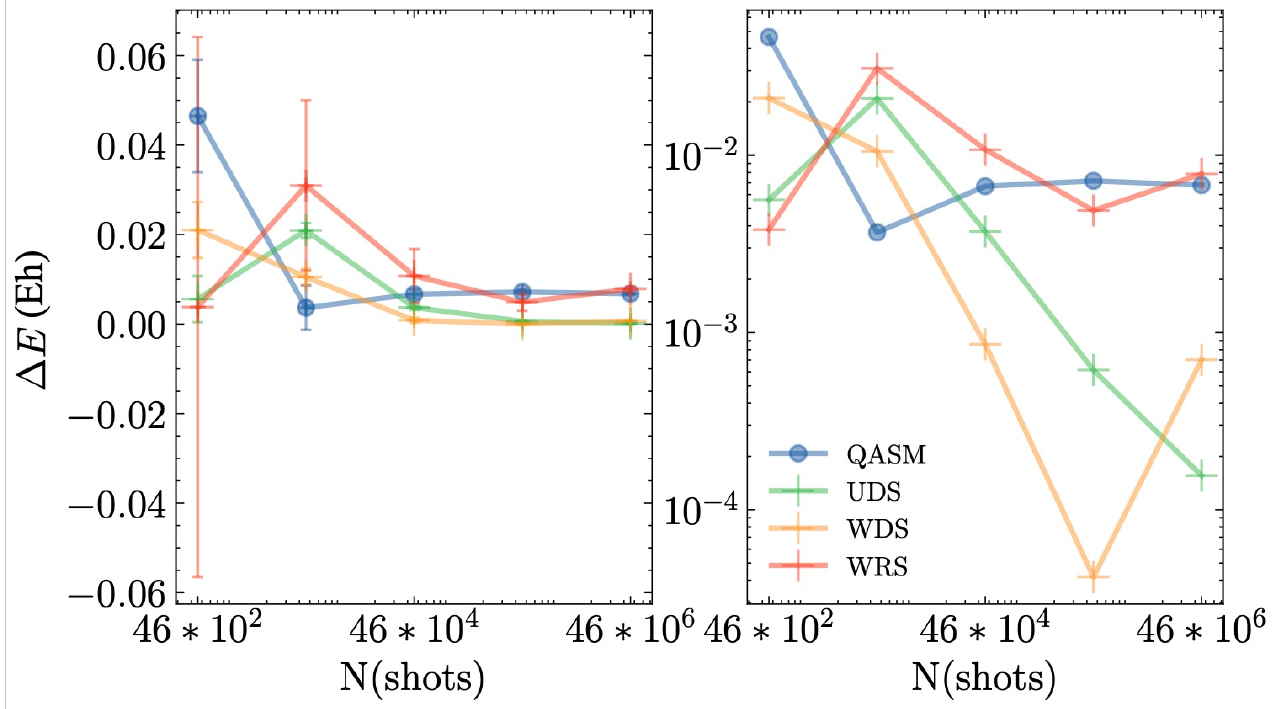}% Here is how to import EPS art
\caption{Shot-frugal technique comparison for the  Hamiltonian operator with the H$_2$ system at equilibrium and with the 6-31G basis set.\label{fig:os_h}}
\end{figure}

\begin{figure}[h!]
\includegraphics[scale=0.65]{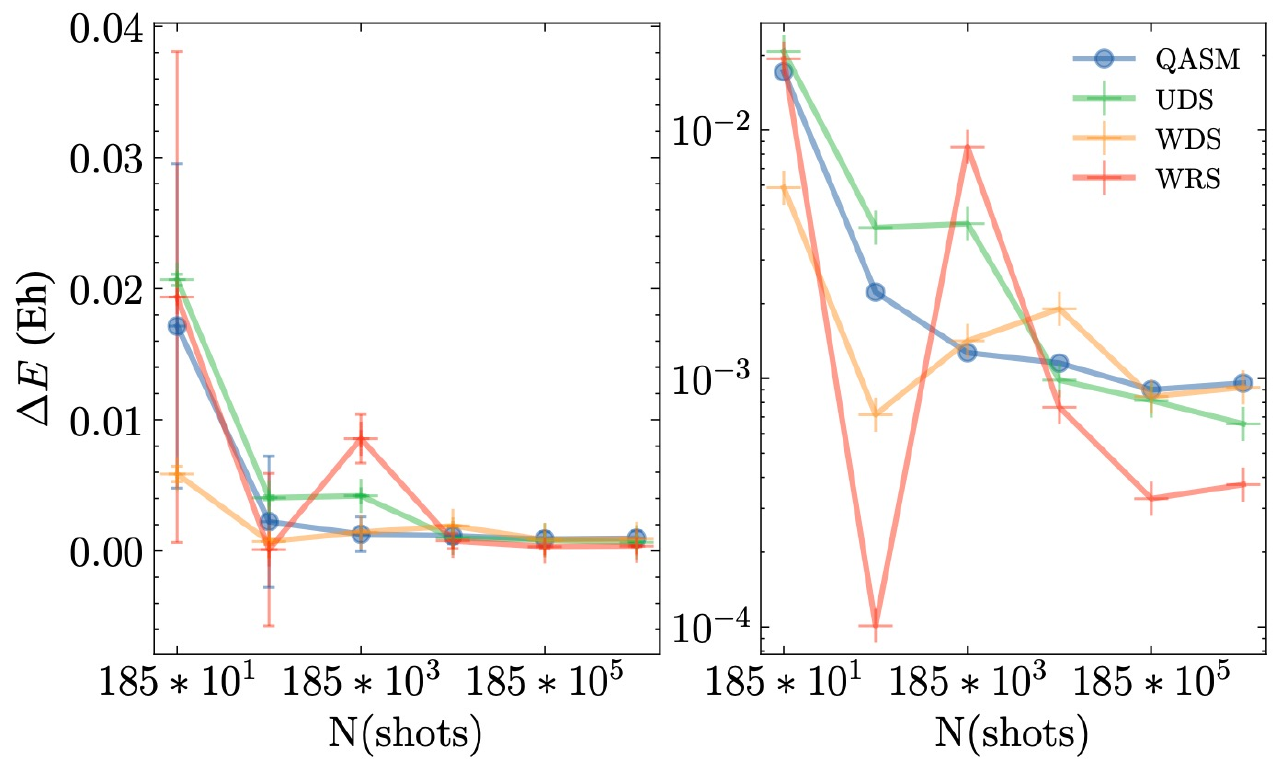}
\caption{Shot-frugal technique comparison for the $J(\vec{\alpha},\vec{\lambda})^{\dagger}HJ(\vec{\alpha},\vec{\lambda})$ operator with the H$_2$ system at equilibrium and 6-31G basis set.\label{fig:os_jhj}}
\end{figure}

We explore three techniques: uniform deterministic sampling (UDS), weighted deterministic sampling (WDS), and weighted random sampling (WRS) and shows results with both the Hamiltonian and $J(\vec{\alpha},\vec{\lambda})^{\dagger}HJ(\vec{\alpha},\vec{\lambda})$ operators respectively in Figure \ref{fig:os_h} and \ref{fig:os_jhj}. 

In UDS, the number of total shots ($s_{tot}$) is divided equally among $N$ terms. In WDS, the number of shots is allocated deterministically proportional to the magnitude of the coefficients. In WRS, the number of shots is sampled from a multinomial probability distribution of which the probability ($p_i$) of measuring a given $\hat{P}_i$ is proportional to the magnitude of the coefficient: $p_i= \frac{|c_i|}{M}$, where $c_i$ is the coefficient and $M$ is the sum of the coefficient. The variances can be expressed as the following:
\begin{align}
\operatorname{Var}(\widehat{E_{UDS}})&=\frac{N}{s_{{tot }}} \sum_{i=1}^N|c_i|^2 \sigma_i^2\\
\operatorname{Var}(\widehat{E_{WDS}})&=\frac{M}{s_{{tot }}} \sum_{i=1}^N|c_i| \sigma_i^2\\
\operatorname{Var}(\widehat{E_{WRS}})&= \frac{M}{s_{\mathrm{tot}}} \sum_{i=1}^N|c_i| \sigma_i^2\notag\\&+\frac{s_{{rand }} M}{s_{{tot }}^2} \sum_{i=1}^N|c_i|\langle \hat{P}_i\rangle^2-\frac{s_{{rand }}\langle \hat{O}\rangle^2}{s_{{tot }}^2}
\end{align}
where $\sigma_i$ is $\langle \hat{P}_i^2\rangle-\langle \hat{P}_i\rangle^2$,
$s_{rand}$ is the random number of shots drawn from the multinomial distribution.

To showcase the shot-frugal techniques, we start with a converged VQE calculation and nuVQE calculation with UCCSD ansatz for the H$_2$ system and 6-31G basis set, and measure the expectation value of the Hamiltonian and $J(\vec{\alpha},\vec{\lambda})^{\dagger}HJ(\vec{\alpha},\vec{\lambda})$ respectively. We use the statevector to calculate the two expectation values exactly. For each method (no shot-frugal, UDS, WDS, WRS), we run 5 independent calculations with the QASM simulation from Qiskit BasicAer module. We calculate the sample mean and take the energy difference from the exact value in Figure \ref{fig:os_h} and \ref{fig:os_jhj}. We also calculate the standard error as $ SEM =\frac{1}{\sqrt{n}}\sqrt{\frac{1}{n}\sum Var_i}$ where $n$ is the number of runs and plot as error bars in the left panels of the two figures. 

As Figure \ref{fig:os_h} right panel shows, all three shot-frugal techniques achieve a lower error with a small number of shots ($10^2$), and UDS and WDS achieve 1-2 orders of magnitude higher accuracy compared to direct QASM measurement with $10^4 - 10^6$ number of shots. It might require the QASM measurements to go even further with the number of shots, compared to the aforementioned two techniques, to obtain a similar error magnitude. Similarly in  Figure \ref{fig:os_jhj} right panel, for operator $J(\vec{\alpha},\vec{\lambda})^{\dagger}HJ(\vec{\alpha},\vec{\lambda})$, we observe that WDS can achieve higher accuracy with low magnitude of shots ($10^1$), and WDS and WRS reaches 1-2 magnitudes higher of accuracy with $10^2$ of shots. WRS shows a more significant advantage with an increased number of shots $10^4 -10^6$. Although not one shot-frugal technique is shown to \textit{always} outperform direct QASM measurements, we see that it oftentimes is more accurate when having extremely limited of shots, and an overall smaller error magnitude with more shots, which would require direct QASM measurement even more number of magnitude of shots to achieve. Notice our results are limited to 5 runs, and the performance of the shot-frugal techniques can change when we have more experiments. It serves as a demonstration that when measuring $J(\vec{\alpha},\vec{\lambda})^{\dagger}HJ(\vec{\alpha},\vec{\lambda})$, one might require less number of shots with the help of shot-frugal techniques. 

\subsubsection{Shot-frugal techniques with qubit-wise grouped Paulis}
To mimic hardware behavior, we further explore the aforementioned three shot-frugal techniques (UDS, WDS, WRS) with QWC grouped Paulis for the $J(\vec{\alpha},\vec{\lambda})^{\dagger}HJ(\vec{\alpha},\vec{\lambda})$ operator. %Hamiltonian operator results are in Appendix B\ref{sec:shot_frugal_qwc_h}.

\begin{figure}[h!]
\includegraphics[scale=0.75]{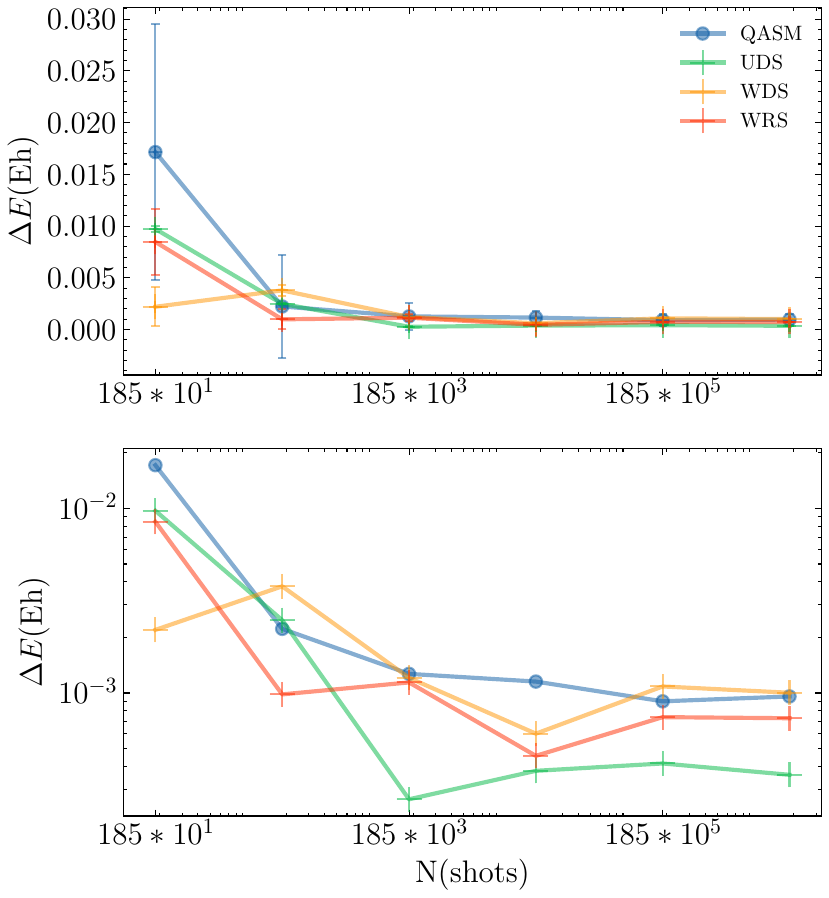}
\caption{Shot-frugal technique with qubit-wise grouped Paulis for the $J(\vec{\alpha},\vec{\lambda})^{\dagger}HJ(\vec{\alpha},\vec{\lambda})$ %operator. 
operator with the H$_2$ system at equilibrium and with the 6-31G basis set.\label{fig:qwc_os_jhj}}
\end{figure}
From Fig. \ref{fig:qwc_os_jhj}, we see that with a small magnitude of shots ($10^1$), all three shot-frugal techniques report more accurate energies compared to QASM. WRS seems to outperform QASM measurements at all number of shots, whereas UDS reports a smaller magnitude of errors with $10^3$ and above magnitude of shots. WDS seems to perform similarly to QASM with $10^1$ and $10^4$ magnitudes being the only two points to report a more accurate energy value compared to QASM. With results in Figure \ref{fig:qwc_os_jhj}, we illustrate that one can further combine the mitigation techniques and mitigate measurement costs for the LAS-nuVQE method. 

\subsection{Scalability}
\QW{We have explored and provided strategies to optimize the LAS-nuVQE algorithms for shorter runtime (by mitigation of measurement costs) and achieving accurate energies with affordable quantum resources.}
In this section, we further provide resource comparison for each methodological development in this work in the form of wall clock time~\cite{Sung_2020}. Specifically, we compare the time required for one iteration of \QW{VQE for vanilla HEA (HEA in the figure), HEA with nuVQE (nuVQE in the figure), and HEA with nuVQE and measurement mitigation}. We follow a similar formalization for walk clock estimation~\cite{Sung_2020}, that time needed for one query can be described by 
$T_{prepare}$ + $T_{sample}$ + $T_{switch}$ + $T_{cloud}$. $T_{prepare}$ is an add-on here to incorporate gate costs and it is the time it takes to initialize a circuit. $T_{sample}$ describes the time for sampling circuits on quantum processors. $T_{switch}$ is the time for the overhead in switching between different circuits, and $T_{cloud}$ describes latency in communication.

For numerical estimations, $T_{prepare}=g * p$, where $g$ is the total number of gates and $p$ is the rate of executing gates. $T_{sample} = M/s$, where M is the number of measurements and $s$ is the sampling rate. $T_{switch} = r * c $, which $r$ represents the overhead in readying and $c$ is the number of diffent circuits. Lastly, $T_{cloud} = l * c/b$, where $l$ is the round-trip time for network and $b$ is the number of circuits sent. For superconducting platform, we set $s = 10^{5}Hz$, $r = 0.1s$, and $l = 4.0s$, and average gate execution time to be $p =660ns$~\cite{abughanem2024ibmquantumcomputersevolution}. For trapped ion/neutral atom platforms, we combine $T_{prepare}$ and $T_{sample}$ to $M/s$ and estimate the shot acquisition time to 200ms ($s=5Hz$), which includes the time for circuit initialization, circuit execution, and one single measurement~\cite{Menickelly2023latency}. The overhead in reading, $r = 0.025s$~\cite{Menickelly2023latency}, and we assume the same $l = 4.0s$.

To carry out calculations for any chemical systems, knowing the number of layers or parameters needed for any HEA to achieve chemical accuracy is almost impossible. Therefore, we assumethe number of layers needed equals to the number of qubits: $N_{l} = N_{q}$. For HEAs, the single qubit gates scale as $N{q}*(1+N_{l})$. Since nuVQE achieves chemical accuracy faster than vanilla HEA-only VQEs, we assume that one Jastrow parameter, which scales as $\frac{1}{2}N_q*(N_q +1)$, can save one SQG. Based on the performance of shot frugal techniques, we assume that the number of shots needed is 2 orders of magnitude less.
\begin{figure}
    \centering
    \includegraphics[scale=0.8]{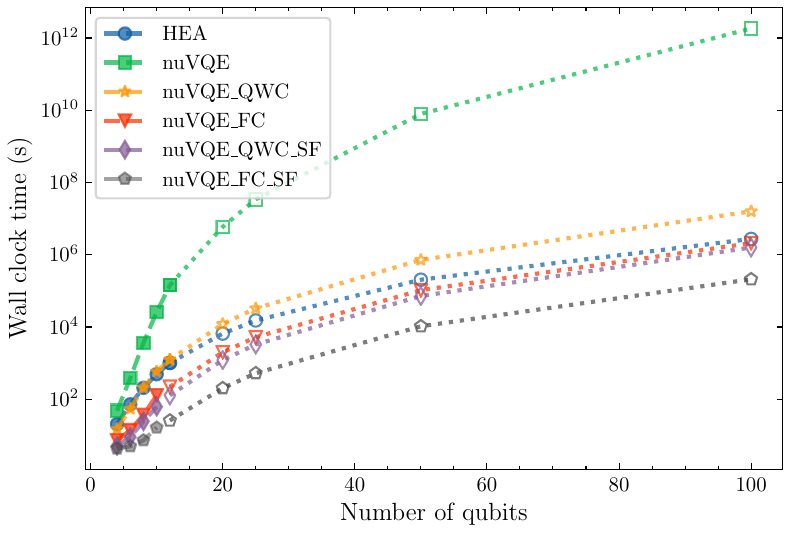}
    \caption{Wall clock estimation for one query for each method on a superconducting platform for hydrogen chains. Hollow markers indicate the extrapolated points.}
    \label{fig:scale-superconduct}
\end{figure}
\begin{figure}
    \centering
    \includegraphics[scale=0.8]{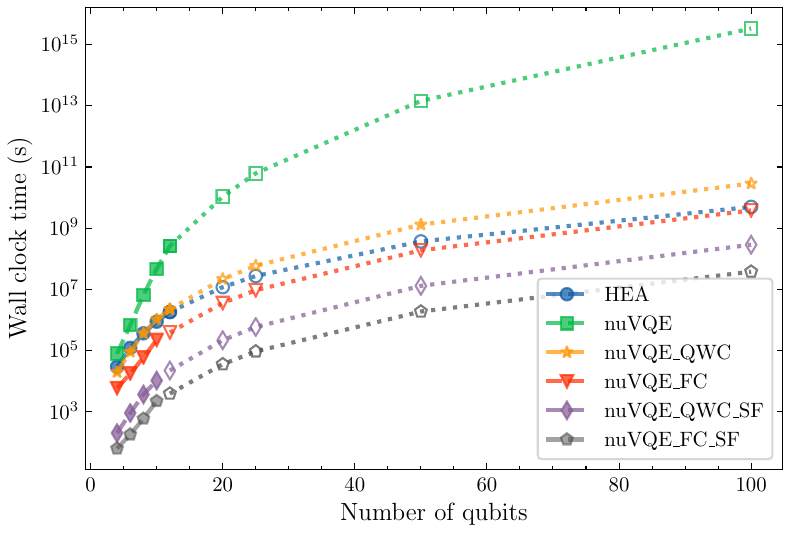}
    \caption{Wall clock estimation for one query for each method on a neutral atom/trapped ion platform for hydrogen chains. Hollow markers indicate the extrapolated points.}
    \label{fig:scale-tina}
\end{figure} 
We first obtained the number of Paulis and grouped Paulis for hydrogen chains H$_2$, H$_3$, H$_4$, H$_5$, H$_6$ (4, 6, 8, 10, 12 qubits respectively). This number (hereafter $h$) also represents the number of unique circuits for the measurements. We then fit \(\ln(h)\) vs. \(\ln(N_q)\) to a straight line then derive constants $a$ and $p$ such that \(h(N_q)\approx a\,n^{p}\). We then use this relationship to extrapolate \(h\) at larger values of \(N_q\), specifically 20, 25, 50, and 100 qubits. We specifically compare six methods: `HEA' represents the well-known vanilla VQE with HEA; `nuVQE' means HEA with VQE; `QWC' indicates qubit-wise grouping is utilized; `FC' means fully-commuting grouping is used; and `SF' is short for shot-frugal.

As Fig. \ref{fig:scale-superconduct} and Fig. \ref{fig:scale-tina} both show, untreated nuVQE quickly becomes unrealistic as the number of qubits increases. At 100 qubits, one query of nuVQE will take about 31,688 years for the superconducting platform and 31,688,764 years for the natural atom/trapped ion platform. HEA will take less than 280 hours on superconducting platforms and 32 years on trapped ion/neutral atom platform. nuVQE with shot-frugal techniques and grouped Paulis can further reduce the time costs down to $<28$ hours on superconducting platforms. In cases where fully-commuting grouping is intractable, qubit-wise grouping and shot-frugal techniques for nuVQE still give a clock time scaling less than HEA, showing its promise in running 100-qubit calculations with realistic wall clock and resources. 

\section{IV. Conclusion \label{Sec:IV}}
In this paper, we have explored the nuVQE method with LASSCF as the initial state. LAS-nuVQE has shown that it can recover interfragment correlations and achieve chemical accuracies with limited quantum resources ($<$70 total gates) for small systems like H$_4$ and C$_4$H$_4$. \QW{To carry out meaningful multireference chemical calculations in the future, we first noted the pitfall of HEA and devised the addition of penalty terms to allow for spin-conversation in the ansatz.} Our results have shown that we can correctly preserve the spin and it allows one to study more complex chemical properties with HEA or HEA-related methods. \QW{Then, to optimize the method further to enable large chemical calculations, we explored Pauli grouping and shot-frugal techniques}. We illustrated that one can significantly reduce measurement costs with grouping and shot-frugal techniques to report a more accurate energy expectation value. Additionally, we provided resource estimation in the form of wall clock time for the methods we explored, showing that with measurement mitigation techniques, LAS-nuVQE remains a competitive candidate for tackling chemical calculations with quantum algorithms. 

\QW{There are several future directions for further investigation in this area. For instance, it would be valuable to explore methods of loading the LASSCF wave function in a way that strictly preserves spin symmetry, thereby maintaining accurate electronic state representations throughout variational optimization. Additionally, examining alternative connectivity patterns for HEA could potential provide deeper insights into achieving faster convergence or improved accuracy by taking advantage of the available qubit connectivity in near-term quantum devices. Another direction involves adopting the local unitary cluter Jastrow (LUCJ) ansatz for capturing electron correlation as another post-LASSCF method. LUCJ is formulated as an approximation to the general unitary cluster Jastrow (UCJ) ansatz, it retains a strong chemical motivation and may offer a cost-effective compromise between accuracy and resource requirements. Furthermore, investigating the performance of LAS-nuVQE with noise models and its performance when combined with error mitigation techniques could suggest whether it exhibits robust noise-resilient features when compared with more conventional ansatz like UCCSD. Such study would be helpful to understand the potential of LAS-nuVQE for practical quantum simulations in NISQ era for reliable electronic structure calculations.}
\appendix
\section{Appendix A: Hardware-efficient ansatz}\label{sec:appendixHEA}
Here we show the HEA with three different connectivities that we explored in this work.
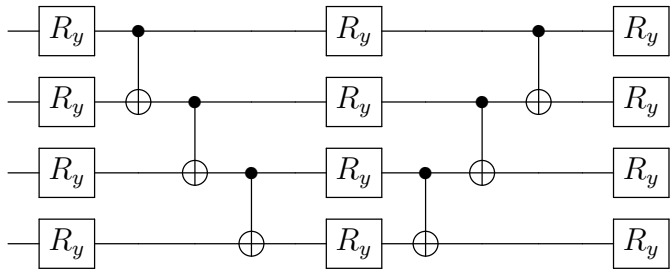
\begin{figure}[h!]
    \centering
    \Qcircuit @C=1em @R=.7em {
        & \gate{R_y} & \ctrl{1} & \qw & \qw & \qw & \gate{R_y} & \qw & \qw & \ctrl{1}   & \qw & \gate{R_y}\\
        & \gate{R_y} & \targ & \ctrl{1} & \qw & \qw &\gate{R_y}& \qw & \ctrl{1} & \targ& \qw  &\gate{R_y} \\
        & \gate{R_y} & \qw & \targ &  \ctrl{1}& \qw & \gate{R_y} & \ctrl{1}  &  \targ & \qw & \qw &\gate{R_y}  \\
        & \gate{R_y} & \qw & \qw & \targ  & \qw &\gate{R_y} & \targ & \qw & \qw  & \qw &\gate{R_y}\\
    }           
    \caption{Cascade connectivity with 4 qubits and 1 layer}
    \label{cascade}
\end{figure}

\begin{figure}[h!]
    \centering
    \Qcircuit @C=1em @R=.7em {
        & \gate{R_y} & \ctrl{1} & \qw & \qw & \qw & \gate{R_y} \\
        & \gate{R_y} & \targ & \ctrl{1} & \qw & \qw &\gate{R_y}  \\
        & \gate{R_y} & \qw & \targ &  \ctrl{1}& \qw & \gate{R_y} \\
        & \gate{R_y} & \qw & \qw & \targ  & \qw &\gate{R_y} \\
    }           
    \caption{Linear connectivity  with 4 qubits and 1 layer}
    \label{linear}
\end{figure}
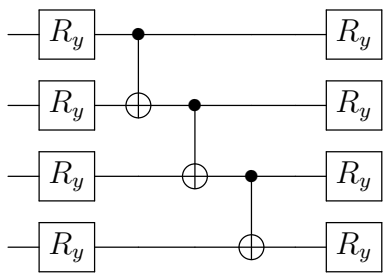

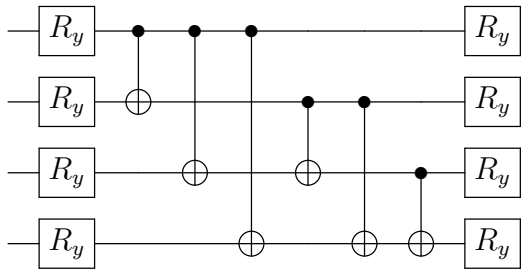
\begin{figure}[h!]
    \centering
    \Qcircuit @C=1em @R=.7em {
    &\gate{R_y}& \ctrl{1}& \ctrl{2}& \ctrl{3}& \qw& \qw& \qw&\gate{R_y}\\
     &\gate{R_y}& \targ& \qw& \qw& \ctrl{1}& \ctrl{2}& \qw&\gate{R_y}\\
     &\gate{R_y}& \qw &\targ & \qw& \targ& \qw& \ctrl{1}&\gate{R_y}\\
     &\gate{R_y}& \qw& \qw& \targ& \qw &\targ &\targ&\gate{R_y}\\
     }
    \caption{Full connectivity  with 4 qubits and 1 layer}
    \label{linear}
\end{figure}
%\section{Appendix B: Shot-frugal techniques with grouping for Hamiltonian }\label{sec:shot_frugal_qwc_h}
%We show the results of shot-frugal techniques with qubit-wise grouped Paulis of the Hamiltonian operator.  
%\begin{figure}[h!]
%\includegraphics[scale=0.7]{Graphs/paper_qwc_os_h.pdf}
%\caption{Shot-frugal technique with qubit-wise grouped Paulis for the Hamiltonian operator with the H$_2$ system at equilibrium and with the 6-31G basis set.\label{fig:qwc_os_h}}
%\end{figure}

\begin{acknowledgement}
The authors thank Dr. Mario Motta, Dr. Kevin Sung, and Dr. Leon Otis for their helpful discussions and feedback on this work. This work was supported by the U.S. Department of Energy, Office of Science, National Quantum Information Science Research Centers and partially by NSF QuBBE Quantum Leap Challenge Institute (NSF OMA-2121044). The authors thank the Research Computing Center at the University of Chicago for the computing resources.

\end{acknowledgement}

\begin{suppinfo}
%\section{Supporting Information}
The supporting information contains the absolute energies for hardware-efficient anstazes comparison, the absolute energies for single-point calculations of two geometries of H$_4$, expectation value measurement data for shot-frugal techniques, and data used for wall clock time calculation. 
%No supporting information is available.
\end{suppinfo}

\section{Data and Code Availability}
The data that support the findings of this study are available within the article and in the supporting information. The code is available at \href{https://github.com/joannaqw/LAS-nuVQE}{https://github.com/joannaqw/LAS-nuVQE}.
\bibliography{achemso-demo}

\end{document}

\tableofcontents

\section{Absolute energies for hardware-efficient ansatzes comparison}
Here we report the absolute energies of Section III Fig. 3 in the manuscript.
\begin{table}[h!]
\begin{tabular}{|l|l|l|l|}
\hline
Number of layers & ry\_linear            & ry\_full              & ry\_cascade           \\ \hline
2                & -2.06706864 & -2.06685570 &   -2.10735438                   \\ \hline
3                & -2.07070057 & -2.06839740 & -2.10746175                    \\ \hline
4                & -2.10561590 & -2.10561545 &  -2.10766108                    \\ \hline
5                & -2.10790506 & -2.10730107   & -2.10813962 \\ \hline
6                & -2.10811983  & -2.10822092   &   \\ \hline
7                & -2.10848992 & -2.10836507   &  \\ \hline
8                & -2.10826548    & -2.10735973  &  \\ \hline
\end{tabular}
\caption{Absolute energies in Hartree of H$_4$ geometry 1, 0 as the initial state}
\end{table}
\begin{table}[h!]
\begin{tabular}{|l|l|l|l|}
\hline
Number of layers & ry\_linear            & ry\_full              & ry\_cascade           \\ \hline
2                & -2.05125139 & -2.05125126 & -2.10387643   \\ \hline
3                & -2.07686156   & -2.07684548 & -2.10747116 \\ \hline
4                & -2.10386723 & -2.10377974 & -2.10840079  \\ \hline
5                & -2.10622805,   & -2.10617792 & -2.10850126   \\ \hline
6                & -2.10850447   & -2.10675286 &                      \\ \hline
7                & -2.10860252  & -2.10855557   &                      \\ \hline
8                & -2.10857281 & -2.10859315   &                      \\ \hline
\end{tabular}
\caption{Absolute energies in Hartree of H$_4$ geometry 1, Hartree-Fock initial state}
\end{table}
\begin{table}[h!]
\begin{tabular}{|l|l|l|l|}
\hline
Number of layers & ry\_linear            & ry\_full              & ry\_cascade           \\ \hline
2                & -2.10263672   & -2.10263608& -2.10593675\\ \hline
3                & -2.07242737 & -2.07565478  & -2.10748325  \\ \hline
4                & -2.10590464  & -2.10589837& -2.10714432 \\ \hline
5                & -2.10465663 & -2.10553014  & -2.10768803 \\ \hline
6                & -2.10751095 & -2.10683177  &                      \\ \hline
7                & -2.10691880 & -2.10808870 &                      \\ \hline
8                & -2.10744628    & -2.10829357  &                      \\ \hline
\end{tabular}
\caption{Absolute energies in Hartree of H$_4$ geometry 1, LASSCF initial state}
\end{table}

\section{Absolute energies for single-point calculations}
The absolute energies of C$_4$H$_4$ are reported in the manuscript, here we provide absolute energies for the two geometries of H$_4$, as graphed in Section III Fig. 4.
\begin{table}[]
\begin{tabular}{|lllll|}
\hline
\multicolumn{5}{|l|}{FCI: -2.10874105} \\ \hline
\multicolumn{1}{|l|}{Number of layers} &
  \multicolumn{1}{l|}{HF-VQE} &
  \multicolumn{1}{l|}{HF-nuVQE} &
  \multicolumn{1}{l|}{LAS-VQE} &
  LAS-nuVQE \\ \hline
\multicolumn{1}{|l|}{2} &
  \multicolumn{1}{l|}{-2.05125139} &
  \multicolumn{1}{l|}{-2.10724625} &
  \multicolumn{1}{l|}{-2.10263673} &
  -2.10729528 \\ \hline
\multicolumn{1}{|l|}{3} &
  \multicolumn{1}{l|}{-2.07686156} &
  \multicolumn{1}{l|}{-2.10865468} &
  \multicolumn{1}{l|}{-2.07242737} &
  -2.10799702 \\ \hline
\multicolumn{1}{|l|}{4} &
  \multicolumn{1}{l|}{-2.10386723} &
  \multicolumn{1}{l|}{-2.10867769} &
  \multicolumn{1}{l|}{-2.10590464} &
  -2.10815127 \\ \hline
\multicolumn{1}{|l|}{5} &
  \multicolumn{1}{l|}{-2.10622805} &
  \multicolumn{1}{l|}{-2.10872266} &
  \multicolumn{1}{l|}{-2.10465663} &
  -2.10827989 \\ \hline
\multicolumn{1}{|l|}{6} &
  \multicolumn{1}{l|}{-2.10850447} &
  \multicolumn{1}{l|}{-2.10870923} &
  \multicolumn{1}{l|}{-2.10751095} &
  -2.10840510 \\ \hline
\end{tabular}
\caption{Absolute energies in Hartree of VQE and nuVQE for HF and LASSCF initial states for molecule H$_4$ geometry 1.}
\end{table}
\begin{table}[]
\begin{tabular}{|lllll|}
\hline
\multicolumn{5}{|l|}{FCI:  -1.99998561} \\ \hline
\multicolumn{1}{|l|}{Number of layers} &
  \multicolumn{1}{l|}{HF-VQE} &
  \multicolumn{1}{l|}{HF-nuVQE} &
  \multicolumn{1}{l|}{LAS-VQE} &
  LAS-nuVQE \\ \hline
\multicolumn{1}{|l|}{2} &
  \multicolumn{1}{l|}{-1.90387411} &
  \multicolumn{1}{l|}{-1.99948967} &
  \multicolumn{1}{l|}{-1.99997558} &
  -1.99998316 \\ \hline
\multicolumn{1}{|l|}{3} &
  \multicolumn{1}{l|}{-1.97901850} &
  \multicolumn{1}{l|}{-1.99998519} &
  \multicolumn{1}{l|}{-1.94411246} &
  -1.99998308 \\ \hline
\multicolumn{1}{|l|}{4} &
  \multicolumn{1}{l|}{-1.98071229} &
  \multicolumn{1}{l|}{-1.99985629} &
  \multicolumn{1}{l|}{-1.99998268} &
  -1.99998359 \\ \hline
\multicolumn{1}{|l|}{5} &
  \multicolumn{1}{l|}{-1.99998295} &
  \multicolumn{1}{l|}{-1.99997854} &
  \multicolumn{1}{l|}{-1.99998224} &
  -1.99998245 \\ \hline
\multicolumn{1}{|l|}{6} &
  \multicolumn{1}{l|}{-1.99998330} &
  \multicolumn{1}{l|}{-1.99998140} &
  \multicolumn{1}{l|}{-1.99998372} &
  -1.99998394\\ \hline
\end{tabular}
\caption{Absolute energies in Hartree of VQE and nuVQE for HF and LASSCF initial states for molecule H$_4$ geometry 2.}
\end{table}

\section{Data for shot-frugal techniques}
We report the 5 independent QASM runs for expectation value measurement of the operators studied in the manuscript respectively for four methods: 1) no technique, 2) UDS, 3) WDS, and 4) WRS. 
\subsection{Shot-frugal techniques on the Hamiltonian operator}
Here we report data represented in Section III Fig. 6, each table includes the 5 independent runs of the expectation value and its measurement variance with QASM.
\begin{table}[!htbp]
\renewcommand{\arraystretch}{1.2}
\begin{tabular}{|lllllll|}
\hline
 &
  \multicolumn{6}{l|}{Exact (statevector):  -1.86677688} \\ \hline
\multicolumn{1}{|l|}{} &
  \multicolumn{1}{l|}{Magnitude of shots} &
  \multicolumn{1}{l|}{$10^2$} &
  \multicolumn{1}{l|}{$10^3$} &
  \multicolumn{1}{l|}{$10^4$} &
  \multicolumn{1}{l|}{$10^5$} &
  $10^6$ \\ \hline
\multicolumn{1}{|l|}{\multirow{2}{*}{1}} &
  \multicolumn{1}{l|}{Energy} &
  \multicolumn{1}{l|}{-1.78558765} &
  \multicolumn{1}{l|}{-1.8872171} &
  \multicolumn{1}{l|}{-1.8585900} &
  \multicolumn{1}{l|}{-1.86735558} &
  -1.86631156 \\ \cline{2-7} 
\multicolumn{1}{|l|}{} &
  \multicolumn{1}{l|}{Variance} &
  \multicolumn{1}{l|}{0.22519875} &
  \multicolumn{1}{l|}{0.18272422} &
  \multicolumn{1}{l|}{0.19148271} &
  \multicolumn{1}{l|}{0.19328438} &
  0.19309999 \\ \hline
\multicolumn{1}{|l|}{\multirow{2}{*}{2}} &
  \multicolumn{1}{l|}{Energy} &
  \multicolumn{1}{l|}{-1.80339591} &
  \multicolumn{1}{l|}{-1.83399411} &
  \multicolumn{1}{l|}{-1.83897492} &
  \multicolumn{1}{l|}{-1.83202028} &
  -1.83265946 \\ \cline{2-7} 
\multicolumn{1}{|l|}{} &
  \multicolumn{1}{l|}{Variance} &
  \multicolumn{1}{l|}{0.15423382} &
  \multicolumn{1}{l|}{0.18653719} &
  \multicolumn{1}{l|}{0.19622481} &
  \multicolumn{1}{l|}{0.19441459} &
  0.19283504 \\ \hline
\multicolumn{1}{|l|}{\multirow{2}{*}{3}} &
  \multicolumn{1}{l|}{Energy} &
  \multicolumn{1}{l|}{-1.91287362} &
  \multicolumn{1}{l|}{-1.88103006} &
  \multicolumn{1}{l|}{-1.86649333} &
  \multicolumn{1}{l|}{-1.86445868} &
  -1.86673549 \\ \cline{2-7} 
\multicolumn{1}{|l|}{} &
  \multicolumn{1}{l|}{Variance} &
  \multicolumn{1}{l|}{0.13281917} &
  \multicolumn{1}{l|}{0.19743969} &
  \multicolumn{1}{l|}{0.19806450} &
  \multicolumn{1}{l|}{0.19445220} &
  0.19301279 \\ \hline
\multicolumn{1}{|l|}{\multirow{2}{*}{4}} &
  \multicolumn{1}{l|}{Energy} &
  \multicolumn{1}{l|}{-1.7742406} &
  \multicolumn{1}{l|}{-1.86958527} &
  \multicolumn{1}{l|}{-1.86549504} &
  \multicolumn{1}{l|}{-1.86607359} &
  -1.86731149 \\ \cline{2-7} 
\multicolumn{1}{|l|}{} &
  \multicolumn{1}{l|}{Variance} &
  \multicolumn{1}{l|}{0.21324145} &
  \multicolumn{1}{l|}{0.18552100} &
  \multicolumn{1}{l|}{0.19760368} &
  \multicolumn{1}{l|}{0.19279340} &
  0.19324865 \\ \hline
\multicolumn{1}{|l|}{\multirow{2}{*}{5}} &
  \multicolumn{1}{l|}{Energy} &
  \multicolumn{1}{l|}{-1.82530655} &
  \multicolumn{1}{l|}{-1.84368991} &
  \multicolumn{1}{l|}{-1.87087965} &
  \multicolumn{1}{l|}{-1.86808048} &
  -1.86694238 \\ \cline{2-7} 
\multicolumn{1}{|l|}{} &
  \multicolumn{1}{l|}{Variance} &
  \multicolumn{1}{l|}{0.15549960} &
  \multicolumn{1}{l|}{0.19859080} &
  \multicolumn{1}{l|}{0.19360509} &
  \multicolumn{1}{l|}{0.19367434} &
  0.19297905 \\ \hline
\end{tabular}
\caption{5 independent QASM runs for Hamiltonian expectation value measurements, no technique utilized. }
\end{table}

% \usepackage{multirow}
\begin{table}[!htbp]
\renewcommand{\arraystretch}{1.2}
\begin{tabular}{|lllllll|}
\hline
\multicolumn{7}{|l|}{Exact (statevector): -1.86677688} \\ \hline
\multicolumn{2}{|l|}{Maginitude of shots} &
  \multicolumn{1}{l|}{$10^2$} &
  \multicolumn{1}{l|}{$10^3$} &
  \multicolumn{1}{l|}{$10^4$} &
  \multicolumn{1}{l|}{$10^5$} &
  \multicolumn{1}{l|}{$10^6$} \\ \hline
\multicolumn{1}{|l|}{\multirow{2}{*}{1}} &
  \multicolumn{1}{l|}{Energy} &
  \multicolumn{1}{l|}{-1.98404006} &
  \multicolumn{1}{l|}{-1.83776212} &
  \multicolumn{1}{l|}{-1.86111514} &
  \multicolumn{1}{l|}{-1.86314465} &
  \multicolumn{1}{l|}{-1.86634820} \\ \cline{2-7} 
\multicolumn{1}{|l|}{} &
  \multicolumn{1}{l|}{Variance} &
  \multicolumn{1}{l|}{2.720384e-5} &
  \multicolumn{1}{l|}{1.401697e-5} &
  \multicolumn{1}{l|}{1.621309e-6} &
  \multicolumn{1}{l|}{1.532435e-7} &
  \multicolumn{1}{l|}{1.564721e-8} \\ \hline
\multicolumn{1}{|l|}{\multirow{2}{*}{2}} &
  \multicolumn{1}{l|}{Energy} &
  \multicolumn{1}{l|}{-1.90367407} &
  \multicolumn{1}{l|}{-1.87870303} &
  \multicolumn{1}{l|}{-1.86455938} &
  \multicolumn{1}{l|}{-1.86792556} &
  \multicolumn{1}{l|}{-1.86712380} \\ \cline{2-7} 
\multicolumn{1}{|l|}{} &
  \multicolumn{1}{l|}{Variance} &
  \multicolumn{1}{l|}{2.897807e-5} &
  \multicolumn{1}{l|}{1.848312e-5} &
  \multicolumn{1}{l|}{1.754431e-6} &
  \multicolumn{1}{l|}{1.427020e-7} &
  \multicolumn{1}{l|}{1.607811e-8} \\ \hline
\multicolumn{1}{|l|}{\multirow{2}{*}{3}} &
  \multicolumn{1}{l|}{Energy} &
  \multicolumn{1}{l|}{-1.83123631} &
  \multicolumn{1}{l|}{-1.82194559} &
  \multicolumn{1}{l|}{-1.85837180} &
  \multicolumn{1}{l|}{-1.87051736} &
  \multicolumn{1}{l|}{1.86696985} \\ \cline{2-7} 
\multicolumn{1}{|l|}{} &
  \multicolumn{1}{l|}{Variance} &
  \multicolumn{1}{l|}{3.670785e-5} &
  \multicolumn{1}{l|}{6.818133e-6} &
  \multicolumn{1}{l|}{1.566606e-6} &
  \multicolumn{1}{l|}{1.525375e-7} &
  \multicolumn{1}{l|}{1.585863e-8} \\ \hline
\multicolumn{1}{|l|}{\multirow{2}{*}{4}} &
  \multicolumn{1}{l|}{Energy} &
  \multicolumn{1}{l|}{-1.75737316} &
  \multicolumn{1}{l|}{-1.82765172} &
  \multicolumn{1}{l|}{-1.86019800} &
  \multicolumn{1}{l|}{-1.86231333} &
  \multicolumn{1}{l|}{-1.86710447} \\ \cline{2-7} 
\multicolumn{1}{|l|}{} &
  \multicolumn{1}{l|}{Variance} &
  \multicolumn{1}{l|}{5.211253e-5} &
  \multicolumn{1}{l|}{1.189108e-5} &
  \multicolumn{1}{l|}{1.605404e-6} &
  \multicolumn{1}{l|}{1.469617e-7} &
  \multicolumn{1}{l|}{1.559111e-8} \\ \hline
\multicolumn{1}{|l|}{\multirow{2}{*}{5}} &
  \multicolumn{1}{l|}{Energy} &
  \multicolumn{1}{l|}{-1.82954989} &
  \multicolumn{1}{l|}{-1.86313943} &
  \multicolumn{1}{l|}{-1.87105868} &
  \multicolumn{1}{l|}{-1.86691182} &
  \multicolumn{1}{l|}{-1.86711700} \\ \cline{2-7} 
\multicolumn{1}{|l|}{} &
  \multicolumn{1}{l|}{Variance} &
  \multicolumn{1}{l|}{0.000527} &
  \multicolumn{1}{l|}{1.547827e-5} &
  \multicolumn{1}{l|}{1.702286e-6} &
  \multicolumn{1}{l|}{1.574338e-7} &
  \multicolumn{1}{l|}{1.568485e-8} \\ \hline
\end{tabular}
\caption{5 independent QASM runs for Hamiltonian expectation value measurements with UDS shot distribution. }
\end{table}

\begin{table}[!htbp]
\renewcommand{\arraystretch}{1.2}
\begin{tabular}{|lllllll|}
\hline
\multicolumn{7}{|l|}{Exact (statevector): -1.86677688} \\ \hline
\multicolumn{2}{|l|}{Maginitude of shots} &
  \multicolumn{1}{l|}{$10^2$} &
  \multicolumn{1}{l|}{$10^3$} &
  \multicolumn{1}{l|}{$10^4$} &
  \multicolumn{1}{l|}{$10^5$} &
  \multicolumn{1}{l|}{$10^6$} \\ \hline
\multicolumn{1}{|l|}{\multirow{2}{*}{1}} &
  \multicolumn{1}{l|}{Energy} &
  \multicolumn{1}{l|}{-1.99160963} &
  \multicolumn{1}{l|}{-1.88836402} &
  \multicolumn{1}{l|}{-1.86788958} &
  \multicolumn{1}{l|}{-1.86655441} &
  \multicolumn{1}{l|}{-1.86592076} \\ \cline{2-7} 
\multicolumn{1}{|l|}{} &
  \multicolumn{1}{l|}{Variance} &
  \multicolumn{1}{l|}{0.000203} &
  \multicolumn{1}{l|}{1.930222e-5} &
  \multicolumn{1}{l|}{1.546408e-6} &
  \multicolumn{1}{l|}{1.547449e-7} &
  \multicolumn{1}{l|}{1.559583e-8} \\ \hline
\multicolumn{1}{|l|}{\multirow{2}{*}{2}} &
  \multicolumn{1}{l|}{Energy} &
  \multicolumn{1}{l|}{-1.84125112} &
  \multicolumn{1}{l|}{-1.90442813} &
  \multicolumn{1}{l|}{-1.86125487} &
  \multicolumn{1}{l|}{-1.86679747} &
  \multicolumn{1}{l|}{-1.86808677} \\ \cline{2-7} 
\multicolumn{1}{|l|}{} &
  \multicolumn{1}{l|}{Variance} &
  \multicolumn{1}{l|}{0.000111} &
  \multicolumn{1}{l|}{1.731245e-5} &
  \multicolumn{1}{l|}{1.567162e-6} &
  \multicolumn{1}{l|}{1.578312e-7} &
  \multicolumn{1}{l|}{1.552005e-8} \\ \hline
\multicolumn{1}{|l|}{\multirow{2}{*}{3}} &
  \multicolumn{1}{l|}{Energy} &
  \multicolumn{1}{l|}{-1.91479475} &
  \multicolumn{1}{l|}{-1.88412841} &
  \multicolumn{1}{l|}{-1.87240571} &
  \multicolumn{1}{l|}{-1.86388963} &
  \multicolumn{1}{l|}{-1.86565262} \\ \cline{2-7} 
\multicolumn{1}{|l|}{} &
  \multicolumn{1}{l|}{Variance} &
  \multicolumn{1}{l|}{0.000226} &
  \multicolumn{1}{l|}{1.396810e-5} &
  \multicolumn{1}{l|}{1.531407e-6} &
  \multicolumn{1}{l|}{1.572822e-7} &
  \multicolumn{1}{l|}{1.572266e-8} \\ \hline
\multicolumn{1}{|l|}{\multirow{2}{*}{4}} &
  \multicolumn{1}{l|}{Energy} &
  \multicolumn{1}{l|}{-1.86554762} &
  \multicolumn{1}{l|}{-1.85965171} &
  \multicolumn{1}{l|}{-1.85995129} &
  \multicolumn{1}{l|}{-1.86910640} &
  \multicolumn{1}{l|}{-1.86445927} \\ \cline{2-7} 
\multicolumn{1}{|l|}{} &
  \multicolumn{1}{l|}{Variance} &
  \multicolumn{1}{l|}{0.00029558} &
  \multicolumn{1}{l|}{1.476847e-5} &
  \multicolumn{1}{l|}{1.667081e-6} &
  \multicolumn{1}{l|}{1.546371e-7} &
  \multicolumn{1}{l|}{1.561786e-8} \\ \hline
\multicolumn{1}{|l|}{\multirow{2}{*}{5}} &
  \multicolumn{1}{l|}{Energy} &
  \multicolumn{1}{l|}{-1.82577435} &
  \multicolumn{1}{l|}{-1.84993780} &
  \multicolumn{1}{l|}{-1.87667489} &
  \multicolumn{1}{l|}{-1.86732812} &
  \multicolumn{1}{l|}{-1.86625611} \\ \cline{2-7} 
\multicolumn{1}{|l|}{} &
  \multicolumn{1}{l|}{Variance} &
  \multicolumn{1}{l|}{0.000133} &
  \multicolumn{1}{l|}{1.361499e-5} &
  \multicolumn{1}{l|}{1.560336e-6} &
  \multicolumn{1}{l|}{1.568367e-7} &
  \multicolumn{1}{l|}{1.558732e-8} \\ \hline
\end{tabular}
\caption{5 independent QASM runs for Hamiltonian expectation value measurements with WDS shot distribution. }
\end{table}
\begin{table}[!htbp]
\renewcommand{\arraystretch}{1.2}
\begin{tabular}{|lllllll|}
\hline
\multicolumn{7}{|l|}{Exact (statevector): -1.86677688} \\ \hline
\multicolumn{2}{|l|}{Maginitude of shots} &
  \multicolumn{1}{l|}{$10^2$} &
  \multicolumn{1}{l|}{$10^3$} &
  \multicolumn{1}{l|}{$10^4$} &
  \multicolumn{1}{l|}{$10^5$} &
  \multicolumn{1}{l|}{$10^6$} \\ \hline
\multicolumn{1}{|l|}{\multirow{2}{*}{1}} &
  \multicolumn{1}{l|}{Energy} &
  \multicolumn{1}{l|}{-1.84360443} &
  \multicolumn{1}{l|}{-1.78188782} &
  \multicolumn{1}{l|}{-1.82838778} &
  \multicolumn{1}{l|}{-1.82668040} &
  \multicolumn{1}{l|}{1.83252410} \\ \cline{2-7} 
\multicolumn{1}{|l|}{} &
  \multicolumn{1}{l|}{Variance} &
  \multicolumn{1}{l|}{0.018260} &
  \multicolumn{1}{l|}{0.001819} &
  \multicolumn{1}{l|}{0.000181} &
  \multicolumn{1}{l|}{1.816530e-5} &
  \multicolumn{1}{l|}{1.815111e-6} \\ \hline
\multicolumn{1}{|l|}{\multirow{2}{*}{2}} &
  \multicolumn{1}{l|}{Energy} &
  \multicolumn{1}{l|}{-1.86660312} &
  \multicolumn{1}{l|}{-1.84460948} &
  \multicolumn{1}{l|}{-1.85708089} &
  \multicolumn{1}{l|}{-1.87157125} &
  \multicolumn{1}{l|}{-1.86549632} \\ \cline{2-7} 
\multicolumn{1}{|l|}{} &
  \multicolumn{1}{l|}{Variance} &
  \multicolumn{1}{l|}{0.018172} &
  \multicolumn{1}{l|}{0.001813} &
  \multicolumn{1}{l|}{0.000181} &
  \multicolumn{1}{l|}{1.813706e-5} &
  \multicolumn{1}{l|}{1.811936e-6} \\ \hline
\multicolumn{1}{|l|}{\multirow{2}{*}{3}} &
  \multicolumn{1}{l|}{Energy} &
  \multicolumn{1}{l|}{-1.86660312} &
  \multicolumn{1}{l|}{-1.84460948} &
  \multicolumn{1}{l|}{-1.85708089} &
  \multicolumn{1}{l|}{-1.87157125} &
  \multicolumn{1}{l|}{-1.86549632} \\ \cline{2-7} 
\multicolumn{1}{|l|}{} &
  \multicolumn{1}{l|}{Variance} &
  \multicolumn{1}{l|}{0.018172} &
  \multicolumn{1}{l|}{0.001813} &
  \multicolumn{1}{l|}{0.000181} &
  \multicolumn{1}{l|}{1.813706e-5} &
  \multicolumn{1}{l|}{1.811936e-6} \\ \hline
\multicolumn{1}{|l|}{\multirow{2}{*}{4}} &
  \multicolumn{1}{l|}{Energy} &
  \multicolumn{1}{l|}{-1.86660312} &
  \multicolumn{1}{l|}{-1.84460948} &
  \multicolumn{1}{l|}{-1.85708089} &
  \multicolumn{1}{l|}{-1.87157125} &
  \multicolumn{1}{l|}{-1.86549632} \\ \cline{2-7} 
\multicolumn{1}{|l|}{} &
  \multicolumn{1}{l|}{Variance} &
  \multicolumn{1}{l|}{0.018172} &
  \multicolumn{1}{l|}{0.001813} &
  \multicolumn{1}{l|}{0.000182} &
  \multicolumn{1}{l|}{1.813705e-5} &
  \multicolumn{1}{l|}{1.811936e-8} \\ \hline
\multicolumn{1}{|l|}{\multirow{2}{*}{5}} &
  \multicolumn{1}{l|}{Energy} &
  \multicolumn{1}{l|}{-1.90950939} &
  \multicolumn{1}{l|}{-1.86337466} &
  \multicolumn{1}{l|}{-1.88043239} &
  \multicolumn{1}{l|}{-1.86807359} &
  \multicolumn{1}{l|}{-1.86560182} \\ \cline{2-7} 
\multicolumn{1}{|l|}{} &
  \multicolumn{1}{l|}{Variance} &
  \multicolumn{1}{l|}{0.018229} &
  \multicolumn{1}{l|}{0.001818} &
  \multicolumn{1}{l|}{0.000181} &
  \multicolumn{1}{l|}{1.813662e-5} &
  \multicolumn{1}{l|}{1.812090e-6} \\ \hline
\end{tabular}
\caption{5 independent QASM runs for Hamiltonian expectation value measurements with WRS shot distribution. }
\end{table}
\FloatBarrier

\subsection{Shot-frugal techniques on the $J(\vec{\alpha},\vec{\lambda})^{\dagger}HJ(\vec{\alpha},\vec{\lambda})$ operator}
Here, we list the data obtained for Section III, Fig. 7. Each table includes the 5 independent runs of the expectation value and its measurement variance with the BasicAer QASM simulator.

\begin{table}[!htbp]
\renewcommand{\arraystretch}{1.2}
\begin{tabular}{|lllllll}
\hline
\multicolumn{1}{|l|}{\multirow{2}{*}{\begin{tabular}[c]{@{}l@{}}Magnitude\\  of shots\end{tabular}}} &
  \multicolumn{2}{l|}{1} &
  \multicolumn{2}{l|}{2} &
  \multicolumn{2}{l|}{3} \\ \cline{2-7} 
\multicolumn{1}{|l|}{} &
  \multicolumn{1}{l|}{Energy} &
  \multicolumn{1}{l|}{Variance} &
  \multicolumn{1}{l|}{Energy} &
  \multicolumn{1}{l|}{Variance} &
  \multicolumn{1}{l|}{Energy} &
  \multicolumn{1}{l|}{Variance} \\ \hline
\multicolumn{1}{|l|}{$10^1$} &
  \multicolumn{1}{l|}{-0.26783049} &
  \multicolumn{1}{l|}{0.011215} &
  \multicolumn{1}{l|}{-0.22829713} &
  \multicolumn{1}{l|}{0.001559} &
  \multicolumn{1}{l|}{-0.2097775} &
  \multicolumn{1}{l|}{0.006377} \\ \hline
\multicolumn{1}{|l|}{$10^2$} &
  \multicolumn{1}{l|}{-0.22832354} &
  \multicolumn{1}{l|}{0.011215} &
  \multicolumn{1}{l|}{-0.25836027} &
  \multicolumn{1}{l|}{0.008911} &
  \multicolumn{1}{l|}{-0.2498289} &
  \multicolumn{1}{l|}{0.006994} \\ \hline
\multicolumn{1}{|l|}{$10^3$} &
  \multicolumn{1}{l|}{-0.24761573} &
  \multicolumn{1}{l|}{0.008851} &
  \multicolumn{1}{l|}{-0.24408648} &
  \multicolumn{1}{l|}{0.009165} &
  \multicolumn{1}{l|}{-0.24056883} &
  \multicolumn{1}{l|}{0.010496} \\ \hline
\multicolumn{1}{|l|}{$10^4$} &
  \multicolumn{1}{l|}{-0.24396113} &
  \multicolumn{1}{l|}{0.009254} &
  \multicolumn{1}{l|}{-0.24394107} &
  \multicolumn{1}{l|}{0.009477} &
  \multicolumn{1}{l|}{-0.24235580} &
  \multicolumn{1}{l|}{0.009503} \\ \hline
\multicolumn{1}{|l|}{$10^5$} &
  \multicolumn{1}{l|}{-0.24428592} &
  \multicolumn{1}{l|}{0.009652} &
  \multicolumn{1}{l|}{-0.24446784} &
  \multicolumn{1}{l|}{0.009668} &
  \multicolumn{1}{l|}{-0.24273584} &
  \multicolumn{1}{l|}{0.009654} \\ \hline
\multicolumn{1}{|l|}{$10^6$} &
  \multicolumn{1}{l|}{-0.24441654} &
  \multicolumn{1}{l|}{0.009674} &
  \multicolumn{1}{l|}{-0.24407784} &
  \multicolumn{1}{l|}{0.009623} &
  \multicolumn{1}{l|}{-0.24258552} &
  \multicolumn{1}{l|}{0.009653} \\ \hline
\multicolumn{7}{|l|}{Exact (statevector): -0.24424998} \\ \hline
\multicolumn{1}{|l|}{} &
  \multicolumn{2}{l|}{4} &
  \multicolumn{2}{l|}{5} &
   &
   \\ \cline{1-5}
\multicolumn{1}{|l|}{$10^1$} &
  \multicolumn{1}{l|}{-0.20818982} &
  \multicolumn{1}{l|}{0.007315} &
  \multicolumn{1}{l|}{-0.22140999} &
  \multicolumn{1}{l|}{0.024659} &
   &
   \\ \cline{1-5}
\multicolumn{1}{|l|}{$10^2$} &
  \multicolumn{1}{l|}{-0.24083003} &
  \multicolumn{1}{l|}{0.011056} &
  \multicolumn{1}{l|}{-0.23277326} &
  \multicolumn{1}{l|}{0.009866} &
   &
   \\ \cline{1-5}
\multicolumn{1}{|l|}{$10^3$} &
  \multicolumn{1}{l|}{-0.24211678} &
  \multicolumn{1}{l|}{0.009327} &
  \multicolumn{1}{l|}{-0.24051326} &
  \multicolumn{1}{l|}{0.008569} &
   &
   \\ \cline{1-5}
\multicolumn{1}{|l|}{$10^4$} &
  \multicolumn{1}{l|}{-0.24243907} &
  \multicolumn{1}{l|}{0.009618} &
  \multicolumn{1}{l|}{-0.24278608} &
  \multicolumn{1}{l|}{0.009919} &
   &
   \\ \cline{1-5}
\multicolumn{1}{|l|}{$10^5$} &
  \multicolumn{1}{l|}{-0.24269029} &
  \multicolumn{1}{l|}{0.009617} &
  \multicolumn{1}{l|}{-0.24256283} &
  \multicolumn{1}{l|}{0.009614} &
   &
   \\ \cline{1-5}
\multicolumn{1}{|l|}{$10^6$} &
  \multicolumn{1}{l|}{-0.24273017} &
  \multicolumn{1}{l|}{0.009644} &
  \multicolumn{1}{l|}{-0.24265010} &
  \multicolumn{1}{l|}{0.009635} &
   &
   \\ \cline{1-5}
\end{tabular}
\caption{5 independent QASM runs for $J(\vec{\alpha},\vec{\lambda})^{\dagger}HJ(\vec{\alpha},\vec{\lambda})$  expectation value measurements, no techniques utilized. \label{tab:jhj_nosf} }
\end{table}

\begin{table}[!htbp]
\renewcommand{\arraystretch}{1.2}
\begin{tabular}{|lllllll}
\hline
\multicolumn{1}{|l|}{\multirow{2}{*}{\begin{tabular}[c]{@{}l@{}}Magnitude\\  of shots\end{tabular}}} &
  \multicolumn{2}{l|}{1} &
  \multicolumn{2}{l|}{2} &
  \multicolumn{2}{l|}{3} \\ \cline{2-7} 
\multicolumn{1}{|l|}{} &
  \multicolumn{1}{l|}{Energy} &
  \multicolumn{1}{l|}{Variance} &
  \multicolumn{1}{l|}{Energy} &
  \multicolumn{1}{l|}{Variance} &
  \multicolumn{1}{l|}{Energy} &
  \multicolumn{1}{l|}{Variance} \\ \hline
\multicolumn{1}{|l|}{$10^1$} &
  \multicolumn{1}{l|}{-0.25658596} &
  \multicolumn{1}{l|}{2.224143e-6} &
  \multicolumn{1}{l|}{-0.25937450} &
  \multicolumn{1}{l|}{2.136049e-6} &
  \multicolumn{1}{l|}{-0.05486035} &
  \multicolumn{1}{l|}{2.179560e-6} \\ \hline
\multicolumn{1}{|l|}{$10^2$} &
  \multicolumn{1}{l|}{-0.24505828} &
  \multicolumn{1}{l|}{2.065440e-7} &
  \multicolumn{1}{l|}{ -0.26924394} &
  \multicolumn{1}{l|}{2.953045e-7} &
  \multicolumn{1}{l|}{-0.23521470} &
  \multicolumn{1}{l|}{ 5.233228e-7} \\ \hline
\multicolumn{1}{|l|}{$10^3$} &
  \multicolumn{1}{l|}{-0.23667217} &
  \multicolumn{1}{l|}{1.245197e-8} &
  \multicolumn{1}{l|}{-0.26176056} &
  \multicolumn{1}{l|}{2.538672e-8} &
  \multicolumn{1}{l|}{-0.19999493} &
  \multicolumn{1}{l|}{2.956747e-8} \\ \hline
\multicolumn{1}{|l|}{$10^4$} &
  \multicolumn{1}{l|}{-0.24691887} &
  \multicolumn{1}{l|}{2.410070e-9} &
  \multicolumn{1}{l|}{-0.24962457} &
  \multicolumn{1}{l|}{2.163757e-9} &
  \multicolumn{1}{l|}{-0.24281207} &
  \multicolumn{1}{l|}{2.313636e-9} \\ \hline
\multicolumn{1}{|l|}{$10^5$} &
  \multicolumn{1}{l|}{-0.24400178} &
  \multicolumn{1}{l|}{2.202132e-10} &
  \multicolumn{1}{l|}{-0.24256921} &
  \multicolumn{1}{l|}{2.064697e-10} &
  \multicolumn{1}{l|}{-0.24397887} &
  \multicolumn{1}{l|}{2.035187e-10} \\ \hline
\multicolumn{1}{|l|}{$10^6$} &
  \multicolumn{1}{l|}{-0.24390079} &
  \multicolumn{1}{l|}{2.172223e-11} &
  \multicolumn{1}{l|}{-0.24435455} &
  \multicolumn{1}{l|}{2.161833e-11} &
  \multicolumn{1}{l|}{-0.24390583} &
  \multicolumn{1}{l|}{2.103260e-11} \\ \hline
\multicolumn{7}{|l|}{Exact (statevector): -0.24424998} \\ \hline
\multicolumn{1}{|l|}{} &
  \multicolumn{2}{l|}{4} &
  \multicolumn{2}{l|}{5} &
   &
   \\ \cline{1-5}
\multicolumn{1}{|l|}{$10^1$} &
  \multicolumn{1}{l|}{-0.29513939} &
  \multicolumn{1}{l|}{2.643755e-6} &
  \multicolumn{1}{l|}{-0.25187272} &
  \multicolumn{1}{l|}{1.381210e-6} &
   &
   \\ \cline{1-5}
\multicolumn{1}{|l|}{$10^2$} &
  \multicolumn{1}{l|}{-0.20184286} &
  \multicolumn{1}{l|}{4.792424e-7} &
  \multicolumn{1}{l|}{-0.24964845} &
  \multicolumn{1}{l|}{7.545063e-7} &
   &
   \\ \cline{1-5}
\multicolumn{1}{|l|}{$10^3$} &
  \multicolumn{1}{l|}{-0.26622731} &
  \multicolumn{1}{l|}{2.278954e-8} &
  \multicolumn{1}{l|}{-0.23560167} &
  \multicolumn{1}{l|}{2.502439e-8} &
   &
   \\ \cline{1-5}
\multicolumn{1}{|l|}{$10^4$} &
  \multicolumn{1}{l|}{-0.24671273} &
  \multicolumn{1}{l|}{2.008753e-9} &
  \multicolumn{1}{l|}{-0.24010434} &
  \multicolumn{1}{l|}{2.425753e-9} &
   &
   \\ \cline{1-5}
\multicolumn{1}{|l|}{$10^5$} &
  \multicolumn{1}{l|}{-0.24357697} &
  \multicolumn{1}{l|}{2.120032e-10} &
  \multicolumn{1}{l|}{-0.24304779} &
  \multicolumn{1}{l|}{2.157948e-10} &
   &
   \\ \cline{1-5}
\multicolumn{1}{|l|}{$10^6$} &
  \multicolumn{1}{l|}{-0.24325343} &
  \multicolumn{1}{l|}{2.114531e-11} &
  \multicolumn{1}{l|}{-0.24255250} &
  \multicolumn{1}{l|}{2.144869e-11} &
   &
   \\ \cline{1-5}
\end{tabular}
\caption{5 independent QASM runs for $J(\vec{\alpha},\vec{\lambda})^{\dagger}HJ(\vec{\alpha},\vec{\lambda})$  expectation value measurements with UDS shot allocation. }
\end{table}

\begin{table}[!htbp]
\renewcommand{\arraystretch}{1.2}
\begin{tabular}{|lllllll}
\hline
\multicolumn{1}{|l|}{\multirow{2}{*}{\begin{tabular}[c]{@{}l@{}}Magnitude\\  of shots\end{tabular}}} &
  \multicolumn{2}{l|}{1} &
  \multicolumn{2}{l|}{2} &
  \multicolumn{2}{l|}{3} \\ \cline{2-7} 
\multicolumn{1}{|l|}{} &
  \multicolumn{1}{l|}{Energy} &
  \multicolumn{1}{l|}{Variance} &
  \multicolumn{1}{l|}{Energy} &
  \multicolumn{1}{l|}{Variance} &
  \multicolumn{1}{l|}{Energy} &
  \multicolumn{1}{l|}{Variance} \\ \hline
\multicolumn{1}{|l|}{$10^1$} &
  \multicolumn{1}{l|}{-0.24504707} &
  \multicolumn{1}{l|}{2.067455e-6} &
  \multicolumn{1}{l|}{-0.24021540} &
  \multicolumn{1}{l|}{1.881334e-6} &
  \multicolumn{1}{l|}{-0.25645378} &
  \multicolumn{1}{l|}{1.195730e-6} \\ \hline
\multicolumn{1}{|l|}{$10^2$} &
  \multicolumn{1}{l|}{-0.24801828} &
  \multicolumn{1}{l|}{2.012343e-7} &
  \multicolumn{1}{l|}{-0.24951621} &
  \multicolumn{1}{l|}{2.676478e-7} &
  \multicolumn{1}{l|}{-0.25614090} &
  \multicolumn{1}{l|}{1.854906e-7} \\ \hline
\multicolumn{1}{|l|}{$10^3$} &
  \multicolumn{1}{l|}{-0.24357328} &
  \multicolumn{1}{l|}{2.276004e-8} &
  \multicolumn{1}{l|}{-0.24366971} &
  \multicolumn{1}{l|}{2.091566e-8} &
  \multicolumn{1}{l|}{-0.24732930} &
  \multicolumn{1}{l|}{2.057348e-8} \\ \hline
\multicolumn{1}{|l|}{$10^4$} &
  \multicolumn{1}{l|}{-0.24217192} &
  \multicolumn{1}{l|}{2.084582e-9} &
  \multicolumn{1}{l|}{-0.24169246} &
  \multicolumn{1}{l|}{2.141552e-9} &
  \multicolumn{1}{l|}{-0.24322619} &
  \multicolumn{1}{l|}{2.203794e-9} \\ \hline
\multicolumn{1}{|l|}{$10^5$} &
  \multicolumn{1}{l|}{-0.24394381} &
  \multicolumn{1}{l|}{2.112938e-10} &
  \multicolumn{1}{l|}{-0.24411084} &
  \multicolumn{1}{l|}{2.145256e-10} &
  \multicolumn{1}{l|}{-0.24315828} &
  \multicolumn{1}{l|}{2.124573e-10} \\ \hline
\multicolumn{1}{|l|}{$10^6$} &
  \multicolumn{1}{l|}{-0.24272780} &
  \multicolumn{1}{l|}{2.126378e-11} &
  \multicolumn{1}{l|}{-0.24440117} &
  \multicolumn{1}{l|}{2.126176e-11} &
  \multicolumn{1}{l|}{-0.24260679} &
  \multicolumn{1}{l|}{2.125984e-11} \\ \hline
\multicolumn{7}{|l|}{Exact (statevector): -0.24424998} \\ \hline
\multicolumn{1}{|l|}{} &
  \multicolumn{2}{l|}{4} &
  \multicolumn{2}{l|}{5} &
   &
   \\ \cline{1-5}
\multicolumn{1}{|l|}{$10^1$} &
  \multicolumn{1}{l|}{-0.26834272} &
  \multicolumn{1}{l|}{3.021152e-6} &
  \multicolumn{1}{l|}{-0.24050390} &
  \multicolumn{1}{l|}{7.602343e-6} &
   &
   \\ \cline{1-5}
\multicolumn{1}{|l|}{$10^2$} &
  \multicolumn{1}{l|}{-0.23183197} &
  \multicolumn{1}{l|}{2.505367e-7} &
  \multicolumn{1}{l|}{-0.23931808} &
  \multicolumn{1}{l|}{2.189503e-7} &
   &
   \\ \cline{1-5}
\multicolumn{1}{|l|}{$10^3$} &
  \multicolumn{1}{l|}{-0.25019134} &
  \multicolumn{1}{l|}{2.096338e-8} &
  \multicolumn{1}{l|}{-0.24357328} &
  \multicolumn{1}{l|}{2.276004e-8} &
   &
   \\ \cline{1-5}
\multicolumn{1}{|l|}{$10^4$} &
  \multicolumn{1}{l|}{-0.24058144} &
  \multicolumn{1}{l|}{2.144731e-9} &
  \multicolumn{1}{l|}{-0.24404133} &
  \multicolumn{1}{l|}{2.146298e-9} &
   &
   \\ \cline{1-5}
\multicolumn{1}{|l|}{$10^5$} &
  \multicolumn{1}{l|}{-0.24342404} &
  \multicolumn{1}{l|}{2.117368e-10} &
  \multicolumn{1}{l|}{-0.24238252} &
  \multicolumn{1}{l|}{2.111990e-10} &
   &
   \\ \cline{1-5}
\multicolumn{1}{|l|}{$10^6$} &
  \multicolumn{1}{l|}{-0.24265743} &
  \multicolumn{1}{l|}{2.128638e-11} &
  \multicolumn{1}{l|}{-0.24425174} &
  \multicolumn{1}{l|}{2.129157e-11} &
   &
   \\ \cline{1-5}
\end{tabular}
\caption{5 independent QASM runs for $J(\vec{\alpha},\vec{\lambda})^{\dagger}HJ(\vec{\alpha},\vec{\lambda})$  expectation value measurements with WDS shot allocation. }
\end{table}

\begin{table}[!htbp]
\renewcommand{\arraystretch}{1.2}
\begin{tabular}{|lllllll}
\hline
\multicolumn{1}{|l|}{\multirow{2}{*}{\begin{tabular}[c]{@{}l@{}}Magnitude\\  of shots\end{tabular}}} &
  \multicolumn{2}{l|}{1} &
  \multicolumn{2}{l|}{2} &
  \multicolumn{2}{l|}{3} \\ \cline{2-7} 
\multicolumn{1}{|l|}{} &
  \multicolumn{1}{l|}{Energy} &
  \multicolumn{1}{l|}{Variance} &
  \multicolumn{1}{l|}{Energy} &
  \multicolumn{1}{l|}{Variance} &
  \multicolumn{1}{l|}{Energy} &
  \multicolumn{1}{l|}{Variance} \\ \hline
\multicolumn{1}{|l|}{$10^1$} &
  \multicolumn{1}{l|}{-0.24769629} &
  \multicolumn{1}{l|}{0.001725} &
  \multicolumn{1}{l|}{-0.26714862} &
  \multicolumn{1}{l|}{0.001783} &
  \multicolumn{1}{l|}{-0.30554121} &
  \multicolumn{1}{l|}{0.001734} \\ \hline
\multicolumn{1}{|l|}{$10^2$} &
  \multicolumn{1}{l|}{-0.25661718} &
  \multicolumn{1}{l|}{0.000167} &
  \multicolumn{1}{l|}{-0.23829303} &
  \multicolumn{1}{l|}{0.000171} &
  \multicolumn{1}{l|}{-0.23979649} &
  \multicolumn{1}{l|}{0.000165} \\ \hline
\multicolumn{1}{|l|}{$10^3$} &
  \multicolumn{1}{l|}{-0.25103890} &
  \multicolumn{1}{l|}{1.707807e-5} &
  \multicolumn{1}{l|}{-0.23206817} &
  \multicolumn{1}{l|}{1.754661e-5} &
  \multicolumn{1}{l|}{-0.23061299} &
  \multicolumn{1}{l|}{1.713962e-5} \\ \hline
\multicolumn{1}{|l|}{$10^4$} &
  \multicolumn{1}{l|}{-0.24490034} &
  \multicolumn{1}{l|}{1.702728e-6} &
  \multicolumn{1}{l|}{-0.24137273} &
  \multicolumn{1}{l|}{1.747172e-6} &
  \multicolumn{1}{l|}{-0.24450226} &
  \multicolumn{1}{l|}{1.671318e-6} \\ \hline
\multicolumn{1}{|l|}{$10^5$} &
  \multicolumn{1}{l|}{-0.24433272} &
  \multicolumn{1}{l|}{1.699095e-7} &
  \multicolumn{1}{l|}{-0.24412259} &
  \multicolumn{1}{l|}{1.732778e-7} &
  \multicolumn{1}{l|}{-0.24449737} &
  \multicolumn{1}{l|}{1.661302e-7} \\ \hline
\multicolumn{1}{|l|}{$10^6$} &
  \multicolumn{1}{l|}{-0.24438514} &
  \multicolumn{1}{l|}{1.715291e-8} &
  \multicolumn{1}{l|}{-0.24282774} &
  \multicolumn{1}{l|}{1.752583e-8} &
  \multicolumn{1}{l|}{-0.24482451} &
  \multicolumn{1}{l|}{1.698406e-8} \\ \hline
\multicolumn{7}{|l|}{Exact (statevector): -0.24424998} \\ \hline
\multicolumn{1}{|l|}{} &
  \multicolumn{2}{l|}{4} &
  \multicolumn{2}{l|}{5} &
   &
   \\ \cline{1-5}
\multicolumn{1}{|l|}{$10^1$} &
  \multicolumn{1}{l|}{-0.24997291} &
  \multicolumn{1}{l|}{0.001784} &
  \multicolumn{1}{l|}{-0.24769629} &
  \multicolumn{1}{l|}{0.001734} &
   &
   \\ \cline{1-5}
\multicolumn{1}{|l|}{$10^2$} &
  \multicolumn{1}{l|}{-0.24725204} &
  \multicolumn{1}{l|}{0.000176} &
  \multicolumn{1}{l|}{-0.23979649} &
  \multicolumn{1}{l|}{0.000165} &
   &
   \\ \cline{1-5}
\multicolumn{1}{|l|}{$10^3$} &
  \multicolumn{1}{l|}{-0.23414178} &
  \multicolumn{1}{l|}{1.731364e-5} &
  \multicolumn{1}{l|}{-0.23061299} &
  \multicolumn{1}{l|}{1.713962e-5} &
   &
   \\ \cline{1-5}
\multicolumn{1}{|l|}{$10^4$} &
  \multicolumn{1}{l|}{-0.24214306} &
  \multicolumn{1}{l|}{1.793580e-6} &
  \multicolumn{1}{l|}{-0.24450226} &
  \multicolumn{1}{l|}{1.671318e-6} &
   &
   \\ \cline{1-5}
\multicolumn{1}{|l|}{$10^5$} &
  \multicolumn{1}{l|}{-0.24215904} &
  \multicolumn{1}{l|}{1.778023e-7} &
  \multicolumn{1}{l|}{-0.24449737} &
  \multicolumn{1}{l|}{1.661302e-7} &
   &
   \\ \cline{1-5}
\multicolumn{1}{|l|}{$10^6$} &
  \multicolumn{1}{l|}{-0.24251697} &
  \multicolumn{1}{l|}{1.772324e-8} &
  \multicolumn{1}{l|}{-0.24482451} &
  \multicolumn{1}{l|}{1.698406e-8} &
   &
   \\ \cline{1-5}
\end{tabular}
\caption{5 independent QASM runs for $J(\vec{\alpha},\vec{\lambda})^{\dagger}HJ(\vec{\alpha},\vec{\lambda})$  expectation value measurements with WRS shot allocation. }
\end{table}
\FloatBarrier
\subsection{Shot-frugal techniques on qubit-wise grouped $J(\vec{\alpha},\vec{\lambda})^{\dagger}HJ(\vec{\alpha},\vec{\lambda})$ operator}
Since QASM simulator already groups the operator with qubit-wise grouping, the $J(\vec{\alpha},\vec{\lambda})^{\dagger}HJ(\vec{\alpha},\vec{\lambda})$ without shot allocation is the same as reported in Table \ref{tab:jhj_nosf}. We list the data obtained for Section III, Fig. 8. Each table includes the 5 independent runs of the expectation value and its measurement variance with the BasicAer QASM simulator.

\begin{table}[!htbp]
\renewcommand{\arraystretch}{1.2}
\begin{tabular}{|lllllll}
\hline
\multicolumn{1}{|l|}{\multirow{2}{*}{\begin{tabular}[c]{@{}l@{}}Magnitude\\  of shots\end{tabular}}} &
  \multicolumn{2}{l|}{1} &
  \multicolumn{2}{l|}{2} &
  \multicolumn{2}{l|}{3} \\ \cline{2-7} 
\multicolumn{1}{|l|}{} &
  \multicolumn{1}{l|}{Energy} &
  \multicolumn{1}{l|}{Variance} &
  \multicolumn{1}{l|}{Energy} &
  \multicolumn{1}{l|}{Variance} &
  \multicolumn{1}{l|}{Energy} &
  \multicolumn{1}{l|}{Variance} \\ \hline
\multicolumn{1}{|l|}{$10^1$} &
  \multicolumn{1}{l|}{-0.26916707} &
  \multicolumn{1}{l|}{7.363553e-6} &
  \multicolumn{1}{l|}{-0.24555546} &
  \multicolumn{1}{l|}{4.590794e-6} &
  \multicolumn{1}{l|}{-0.25607224} &
  \multicolumn{1}{l|}{3.631203e-6} \\ \hline
\multicolumn{1}{|l|}{$10^2$} &
  \multicolumn{1}{l|}{-0.25310528} &
  \multicolumn{1}{l|}{3.530633e-7} &
  \multicolumn{1}{l|}{-0.25912772} &
  \multicolumn{1}{l|}{3.720340e-7} &
  \multicolumn{1}{l|}{-0.24127215} &
  \multicolumn{1}{l|}{3.303953e-7} \\ \hline
\multicolumn{1}{|l|}{$10^3$} &
  \multicolumn{1}{l|}{-0.24458990} &
  \multicolumn{1}{l|}{3.200804e-8} &
  \multicolumn{1}{l|}{-0.24328174} &
  \multicolumn{1}{l|}{2.742056e-8} &
  \multicolumn{1}{l|}{-0.24844089} &
  \multicolumn{1}{l|}{4.033349e-8} \\ \hline
\multicolumn{1}{|l|}{$10^4$} &
  \multicolumn{1}{l|}{-0.24531200} &
  \multicolumn{1}{l|}{3.230277e-9} &
  \multicolumn{1}{l|}{-0.24440638} &
  \multicolumn{1}{l|}{3.160148e-9} &
  \multicolumn{1}{l|}{-0.24343880} &
  \multicolumn{1}{l|}{3.411559e-9} \\ \hline
\multicolumn{1}{|l|}{$10^5$} &
  \multicolumn{1}{l|}{-0.24422893} &
  \multicolumn{1}{l|}{3.285168e-10} &
  \multicolumn{1}{l|}{-0.24408078} &
  \multicolumn{1}{l|}{3.277178e-10} &
  \multicolumn{1}{l|}{-0.24405448} &
  \multicolumn{1}{l|}{3.285908e-10} \\ \hline
\multicolumn{1}{|l|}{$10^6$} &
  \multicolumn{1}{l|}{-0.24429981} &
  \multicolumn{1}{l|}{3.270995e-11} &
  \multicolumn{1}{l|}{-0.24415249} &
  \multicolumn{1}{l|}{3.283142e-11} &
  \multicolumn{1}{l|}{-0.24421493} &
  \multicolumn{1}{l|}{3.275126e-11} \\ \hline
\multicolumn{7}{|l|}{Exact (statevector): -0.24424998} \\ \hline
\multicolumn{1}{|l|}{} &
  \multicolumn{2}{l|}{4} &
  \multicolumn{2}{l|}{5} &
   &
   \\ \cline{1-5}
\multicolumn{1}{|l|}{$10^1$} &
  \multicolumn{1}{l|}{-0.25339090} &
  \multicolumn{1}{l|}{3.724613e-6} &
  \multicolumn{1}{l|}{-0.24563765} &
  \multicolumn{1}{l|}{2.747529e-6} &
   &
   \\ \cline{1-5}
\multicolumn{1}{|l|}{$10^2$} &
  \multicolumn{1}{l|}{-0.24055650} &
  \multicolumn{1}{l|}{6.755313e-7} &
  \multicolumn{1}{l|}{-0.23959154} &
  \multicolumn{1}{l|}{6.676485e-7} &
   &
   \\ \cline{1-5}
\multicolumn{1}{|l|}{$10^3$} &
  \multicolumn{1}{l|}{-0.24470611} &
  \multicolumn{1}{l|}{3.516903e-8} &
  \multicolumn{1}{l|}{-0.24155899} &
  \multicolumn{1}{l|}{3.659503e-8} &
   &
   \\ \cline{1-5}
\multicolumn{1}{|l|}{$10^4$} &
  \multicolumn{1}{l|}{-0.24419408} &
  \multicolumn{1}{l|}{3.368663e-9} &
  \multicolumn{1}{l|}{-0.24200488} &
  \multicolumn{1}{l|}{3.484749e-9} &
   &
   \\ \cline{1-5}
\multicolumn{1}{|l|}{$10^5$} &
  \multicolumn{1}{l|}{-0.24425806} &
  \multicolumn{1}{l|}{3.328597e-10} &
  \multicolumn{1}{l|}{-0.24255141} &
  \multicolumn{1}{l|}{3.357394e-10} &
   &
   \\ \cline{1-5}
\multicolumn{1}{|l|}{$10^6$} &
  \multicolumn{1}{l|}{-0.24425231} &
  \multicolumn{1}{l|}{3.270097e-11} &
  \multicolumn{1}{l|}{-0.24252832} &
  \multicolumn{1}{l|}{3.274726e-11} &
   &
   \\ \cline{1-5}
\end{tabular}
\caption{5 independent QASM runs for qubit-wise grouped $J(\vec{\alpha},\vec{\lambda})^{\dagger}HJ(\vec{\alpha},\vec{\lambda})$  expectation value measurements with UDS shot allocation. }
\end{table}
\begin{table}[!htbp]
\renewcommand{\arraystretch}{1.2}
\begin{tabular}{|lllllll}
\hline
\multicolumn{1}{|l|}{\multirow{2}{*}{\begin{tabular}[c]{@{}l@{}}Magnitude\\  of shots\end{tabular}}} &
  \multicolumn{2}{l|}{1} &
  \multicolumn{2}{l|}{2} &
  \multicolumn{2}{l|}{3} \\ \cline{2-7} 
\multicolumn{1}{|l|}{} &
  \multicolumn{1}{l|}{Energy} &
  \multicolumn{1}{l|}{Variance} &
  \multicolumn{1}{l|}{Energy} &
  \multicolumn{1}{l|}{Variance} &
  \multicolumn{1}{l|}{Energy} &
  \multicolumn{1}{l|}{Variance} \\ \hline
\multicolumn{1}{|l|}{$10^1$} &
  \multicolumn{1}{l|}{-0.22446531} &
  \multicolumn{1}{l|}{1.433433e-6} &
  \multicolumn{1}{l|}{-0.29215694} &
  \multicolumn{1}{l|}{1.382614e-6} &
  \multicolumn{1}{l|}{-0.23985356} &
  \multicolumn{1}{l|}{1.969144e-6} \\ \hline
\multicolumn{1}{|l|}{$10^2$} &
  \multicolumn{1}{l|}{-0.23903883} &
  \multicolumn{1}{l|}{1.515875e-7} &
  \multicolumn{1}{l|}{-0.26075102} &
  \multicolumn{1}{l|}{1.324931e-7} &
  \multicolumn{1}{l|}{-0.24048204} &
  \multicolumn{1}{l|}{1.445904e-7} \\ \hline
\multicolumn{1}{|l|}{$10^3$} &
  \multicolumn{1}{l|}{-0.24375634} &
  \multicolumn{1}{l|}{1.541395e-8} &
  \multicolumn{1}{l|}{-0.24565669} &
  \multicolumn{1}{l|}{1.523015e-8} &
  \multicolumn{1}{l|}{-0.24338516} &
  \multicolumn{1}{l|}{1.545580e-8} \\ \hline
\multicolumn{1}{|l|}{$10^4$} &
  \multicolumn{1}{l|}{-0.24355772} &
  \multicolumn{1}{l|}{1.528247e-9} &
  \multicolumn{1}{l|}{-0.24499725} &
  \multicolumn{1}{l|}{1.5240312e-9} &
  \multicolumn{1}{l|}{-0.24479207} &
  \multicolumn{1}{l|}{1.510858e-9} \\ \hline
\multicolumn{1}{|l|}{$10^5$} &
  \multicolumn{1}{l|}{-0.24251750} &
  \multicolumn{1}{l|}{1.511720e-10} &
  \multicolumn{1}{l|}{-0.24438516} &
  \multicolumn{1}{l|}{1.511663e-10} &
  \multicolumn{1}{l|}{-0.24406198} &
  \multicolumn{1}{l|}{1.518075e-10} \\ \hline
\multicolumn{1}{|l|}{$10^6$} &
  \multicolumn{1}{l|}{-0.24259088} &
  \multicolumn{1}{l|}{1.516073e-11} &
  \multicolumn{1}{l|}{-0.24434816} &
  \multicolumn{1}{l|}{1.517148e-11} &
  \multicolumn{1}{l|}{-0.24416815} &
  \multicolumn{1}{l|}{1.517427e-11} \\ \hline
\multicolumn{7}{|l|}{Exact (statevector): -0.24424998} \\ \hline
\multicolumn{1}{|l|}{} &
  \multicolumn{2}{l|}{4} &
  \multicolumn{2}{l|}{5} &
   &
   \\ \cline{1-5}
\multicolumn{1}{|l|}{$10^1$} &
  \multicolumn{1}{l|}{-0.22622965} &
  \multicolumn{1}{l|}{2.472676e-6} &
  \multicolumn{1}{l|}{-0.22755702} &
  \multicolumn{1}{l|}{1.593173e-6} &
   &
   \\ \cline{1-5}
\multicolumn{1}{|l|}{$10^2$} &
  \multicolumn{1}{l|}{-0.25190976} &
  \multicolumn{1}{l|}{1.491029e-7} &
  \multicolumn{1}{l|}{-0.24798922} &
  \multicolumn{1}{l|}{1.243560e-7} &
   &
   \\ \cline{1-5}
\multicolumn{1}{|l|}{$10^3$} &
  \multicolumn{1}{l|}{-0.24387153} &
  \multicolumn{1}{l|}{1.441248e-8} &
  \multicolumn{1}{l|}{-0.23853692} &
  \multicolumn{1}{l|}{1.607295e-8} &
   &
   \\ \cline{1-5}
\multicolumn{1}{|l|}{$10^4$} &
  \multicolumn{1}{l|}{-0.24241623} &
  \multicolumn{1}{l|}{1.524850e-9} &
  \multicolumn{1}{l|}{-0.24247560} &
  \multicolumn{1}{l|}{1.522755e-9} &
   &
   \\ \cline{1-5}
\multicolumn{1}{|l|}{$10^5$} &
  \multicolumn{1}{l|}{-0.24233724} &
  \multicolumn{1}{l|}{1.516709e-10} &
  \multicolumn{1}{l|}{-0.24252146} &
  \multicolumn{1}{l|}{1.521981e-10} &
   &
   \\ \cline{1-5}
\multicolumn{1}{|l|}{$10^6$} &
  \multicolumn{1}{l|}{-0.24261671} &
  \multicolumn{1}{l|}{1.516682e-11} &
  \multicolumn{1}{l|}{-0.24250990} &
  \multicolumn{1}{l|}{1.516140e-11} &
   &
   \\ \cline{1-5}
\end{tabular}
\caption{5 independent QASM runs for qubit-wise grouped $J(\vec{\alpha},\vec{\lambda})^{\dagger}HJ(\vec{\alpha},\vec{\lambda})$ expectation value measurements with WDS shot allocation. }
\end{table}

\begin{table}[!htbp]
\renewcommand{\arraystretch}{1.2}
\begin{tabular}{|lllllll}
\hline
\multicolumn{1}{|l|}{\multirow{2}{*}{\begin{tabular}[c]{@{}l@{}}Magnitude\\  of shots\end{tabular}}} &
  \multicolumn{2}{l|}{1} &
  \multicolumn{2}{l|}{2} &
  \multicolumn{2}{l|}{3} \\ \cline{2-7} 
\multicolumn{1}{|l|}{} &
  \multicolumn{1}{l|}{Energy} &
  \multicolumn{1}{l|}{Variance} &
  \multicolumn{1}{l|}{Energy} &
  \multicolumn{1}{l|}{Variance} &
  \multicolumn{1}{l|}{Energy} &
  \multicolumn{1}{l|}{Variance} \\ \hline
\multicolumn{1}{|l|}{$10^1$} &
  \multicolumn{1}{l|}{-0.24406655} &
  \multicolumn{1}{l|}{0.005214} &
  \multicolumn{1}{l|}{-0.24644741} &
  \multicolumn{1}{l|}{0.004894} &
  \multicolumn{1}{l|}{-0.24644741} &
  \multicolumn{1}{l|}{0.005165} \\ \hline
\multicolumn{1}{|l|}{$10^2$} &
  \multicolumn{1}{l|}{-0.23878764} &
  \multicolumn{1}{l|}{0.000467} &
  \multicolumn{1}{l|}{-0.24263589} &
  \multicolumn{1}{l|}{0.000470} &
  \multicolumn{1}{l|}{0.24263588} &
  \multicolumn{1}{l|}{0.000407} \\ \hline
\multicolumn{1}{|l|}{$10^3$} &
  \multicolumn{1}{l|}{-0.24552762} &
  \multicolumn{1}{l|}{5.030780e-5} &
  \multicolumn{1}{l|}{-0.24600075} &
  \multicolumn{1}{l|}{5.004213e-5} &
  \multicolumn{1}{l|}{-0.24600075} &
  \multicolumn{1}{l|}{5.004213e-5} \\ \hline
\multicolumn{1}{|l|}{$10^4$} &
  \multicolumn{1}{l|}{-0.24421507} &
  \multicolumn{1}{l|}{7.445429e-6} &
  \multicolumn{1}{l|}{-0.244487180} &
  \multicolumn{1}{l|}{7.398101e-6} &
  \multicolumn{1}{l|}{-0.24448717} &
  \multicolumn{1}{l|}{7.398101e-6} \\ \hline
\multicolumn{1}{|l|}{$10^5$} &
  \multicolumn{1}{l|}{-0.24404305} &
  \multicolumn{1}{l|}{7.571229e-7} &
  \multicolumn{1}{l|}{-0.24406418} &
  \multicolumn{1}{l|}{7.477489e-7} &
  \multicolumn{1}{l|}{-0.24406418} &
  \multicolumn{1}{l|}{7.477488e-7} \\ \hline
\multicolumn{1}{|l|}{$10^6$} &
  \multicolumn{1}{l|}{-0.24419827} &
  \multicolumn{1}{l|}{7.549790e-8} &
  \multicolumn{1}{l|}{-0.24417158} &
  \multicolumn{1}{l|}{7.493353e-8} &
  \multicolumn{1}{l|}{-0.24417158} &
  \multicolumn{1}{l|}{7.493352e-8} \\ \hline
\multicolumn{7}{|l|}{Exact (statevector): -0.24424998} \\ \hline
\multicolumn{1}{|l|}{} &
  \multicolumn{2}{l|}{4} &
  \multicolumn{2}{l|}{5} &
   &
   \\ \cline{1-5}
\multicolumn{1}{|l|}{$10^1$} &
  \multicolumn{1}{l|}{-0.22097762} &
  \multicolumn{1}{l|}{0.005164} &
  \multicolumn{1}{l|}{-0.22097761} &
  \multicolumn{1}{l|}{0.004894} &
   &
   \\ \cline{1-5}
\multicolumn{1}{|l|}{$10^2$} &
  \multicolumn{1}{l|}{-0.24613750} &
  \multicolumn{1}{l|}{0.000461} &
  \multicolumn{1}{l|}{-0.24613750} &
  \multicolumn{1}{l|}{0.000470} &
   &
   \\ \cline{1-5}
\multicolumn{1}{|l|}{$10^3$} &
  \multicolumn{1}{l|}{-0.24471341} &
  \multicolumn{1}{l|}{5.004213e-5} &
  \multicolumn{1}{l|}{-0.24471342} &
  \multicolumn{1}{l|}{5.191929e-5} &
   &
   \\ \cline{1-5}
\multicolumn{1}{|l|}{$10^4$} &
  \multicolumn{1}{l|}{-0.24288869} &
  \multicolumn{1}{l|}{7.418957e-6} &
  \multicolumn{1}{l|}{-0.24288869} &
  \multicolumn{1}{l|}{7.418957e-6} &
   &
   \\ \cline{1-5}
\multicolumn{1}{|l|}{$10^5$} &
  \multicolumn{1}{l|}{-0.24268925} &
  \multicolumn{1}{l|}{7.477489e-7} &
  \multicolumn{1}{l|}{-0.24268926} &
  \multicolumn{1}{l|}{7.571229e-7} &
   &
   \\ \cline{1-5}
\multicolumn{1}{|l|}{$10^6$} &
  \multicolumn{1}{l|}{-0.24252913} &
  \multicolumn{1}{l|}{7.4933525e-8} &
  \multicolumn{1}{l|}{-0.24252913} &
  \multicolumn{1}{l|}{7.525947e-8} &
   &
   \\ \cline{1-5}
\end{tabular}
\caption{5 independent QASM runs for qubit-wise grouped $J(\vec{\alpha},\vec{\lambda})^{\dagger}HJ(\vec{\alpha},\vec{\lambda})$  expectation value measurements with WRS shot allocation. }
\end{table}

\FloatBarrier
\section{Data for wall clock time calculation}
We report the number of Pauli strings here in support of Section III. Fig. 9, 10. 
\begin{table}[h!]
\renewcommand{\arraystretch}{1.4}
\begin{tabular}{|l|l|l|l|}
\hline
Number of qubits & Hamiltonian   & $J(\vec{\alpha},\vec{\lambda})^{\dagger}HJ(\vec{\alpha},\vec{\lambda})$                  & J$J(\vec{\alpha},\vec{\lambda})^{\dagger}J(\vec{\alpha},\vec{\lambda})$             \\ \hline
4                & 15            & 24                  & 16            \\ \hline
6                & 62            & 288                 & 57            \\ \hline
8                & 185           & 3,147                & 163           \\ \hline
10               & 444           & 23,096               & 186           \\ \hline
12               & 919           & 130,346              & 794           \\ \hline
20               & $\sim$5,971    & $\sim$5,373,347       & $\sim$4,548    \\ \hline
25               & $\sim$13,784   & $\sim$31,022,282      & $\sim$10,069   \\ \hline
50               & $\sim$185,374  & $\sim$7,192,628,929    & $\sim$118,873  \\ \hline
100              & $\sim$2,493,041 & $\sim$1,667,637,171,587 & $\sim$1,403,332 \\ \hline
\end{tabular}
\caption{Number of ungrouped Pauli strings for hydrogen chains and extrapolated data for 20, 25, 50, 100 qubits.}
\end{table}

\begin{table}[h!]
\renewcommand{\arraystretch}{1.4}
\begin{tabular}{|l|l|l|l|}
\hline
Number of qubits & Hamiltonian   & $J(\vec{\alpha},\vec{\lambda})^{\dagger}HJ(\vec{\alpha},\vec{\lambda})$                  & J$J(\vec{\alpha},\vec{\lambda})^{\dagger}J(\vec{\alpha},\vec{\lambda})$             \\ \hline
4                & 5            & 9                  & 1           \\ \hline
6                & 13            &45                 & 1            \\ \hline
8                & 46           & 185                & 1           \\ \hline
10               & 82          & 528               & 1           \\ \hline
12               & $\sim$145           & $\sim$1,126              & 1           \\ \hline
20               & $\sim$727    & $\sim$11,003       & 1    \\ \hline
25               & $\sim$1,470   & $\sim$29,776      & 1   \\ \hline
50               & $\sim$13,080  & $\sim$655,833    & 1  \\ \hline
100              & $\sim$116,395 & $\sim$14,444,873 & 1 \\ \hline
\end{tabular}
\caption{Number of qubit-wise grouped Pauli strings for hydrogen chains and extrapolated data for 12, 20, 25, 50, 100 qubits.}
\end{table}
\begin{table}[h!]
\renewcommand{\arraystretch}{1.4}
\begin{tabular}{|l|l|l|l|}
\hline
Number of qubits & Hamiltonian   & $J(\vec{\alpha},\vec{\lambda})^{\dagger}HJ(\vec{\alpha},\vec{\lambda})$                  & J$J(\vec{\alpha},\vec{\lambda})^{\dagger}J(\vec{\alpha},\vec{\lambda})$             \\ \hline
4                & 3           & 2                  & 1           \\ \hline
6                & 6            &8                 & 1            \\ \hline
8                & 9           & 29                & 1           \\ \hline
10               & 28          & 112              & 1           \\ \hline
12               & $\sim$36           & $\sim$197              & 1           \\ \hline
20               & $\sim$146    & $\sim$1,804      & 1    \\ \hline
25               & $\sim$266   & $\sim$4,737      & 1   \\ \hline
50               & $\sim$1,728  & $\sim$95,034    & 1  \\ \hline
100              & $\sim$11,199 & $\sim$1,906,196 & 1 \\ \hline
\end{tabular}
\caption{Number of full-commuting grouped Pauli strings for hydrogen chains and extrapolated data for 12, 20, 25, 50, 100 qubits.}
\end{table}